# Localized Mutual Information Monitoring of Pairwise Associations in Animal Movement

Andrew B. Whetten

**Abstract**

*Advances in satellite imaging and GPS tracking devices have given rise to a new era of remote sensing and geospatial analysis. In environmental science and conservation ecology, biotelemetric data is often high-dimensional, spatially and/or temporally, and functional in nature, meaning that there is an underlying continuity to the biological process of interest. GPS-tracking of animal movement is commonly characterized by irregular time-recording of animal position, and the movement relationships between animals are prone to sudden change. In this paper, I propose a measure of localized mutual information (LMI) to derive a correlation function for monitoring changes in the pairwise association between animal movement trajectories. The properties of the LMI measure are assessed analytically and by simulation under a variety of circumstances. Advantages and disadvantages of the LMI measure are assessed and an alternate measure of LMI is proposed to handle potential disadvantages. The measure of LMI is shown to be an effective tool for detecting shifts in the correlation of animal movements, and seasonal/phasal correlatory structure.*

## 1. Introduction

In the past 2 decades, the rapid increase in the quality of GPS-tracking technology has revolutionized the field of movement ecology [1,2]. The advancements in animal-tracking systems have yielded high-precision and densely recorded datasets of wildlife behavior [3]. The current challenge in the field of movement ecology is the appropriate processing and integration of large volumes of tracked animals and local environmental data into a unified modeling process to improve the quality of population level inferences [4,5,6].

Many models have been proposed in recent years to improve animal trajectory estimation, relationships of animal movements, and home-range and land use estimates for individual animals [4-15]. However, in spite of the exceptional characteristics of these models, there is minimal development of descriptive statistical tools [16, 17, 18, 19]. Descriptive/summary statistics include the widely familiar and elementary topics in statistics and probability such as mean, median, variance, covariance, quantiles, $r^2$, and many others. Additionally, there is rich literature on dissimilarity measures and distance metrics that have paralleled the rise of machine learning modeling [20-23]. Although a seemingly trivial topic in modern animal movement modeling, these descriptive measures are the foundation of any sound modeling process, and the appropriate construction and selection of a measure inevitably provides increased insight into the process of interest [24, 25].

Reflecting on the rapid progression of the field of movement ecology, I propose the following question: *What does correlation mean in animal movement?* The correlation coefficient is a descriptive measure used widely across all branches of statistics. A few household measures of correlation are listed here: Pearson, Spearman, Kendall [26, 27]. In information theory, a common measure of association is referred to as mutual information, which is a quantification of the amount of information obtained from the distribution of a random variable by observing another random variable [28]. Although correlation measures are often abused or "over-interpreted," they can be thought of as a measure of similarity/ dissimilarity between observations or variables in a dataset [24, 29].

The use of information theoretic measures in movement ecology has modestly increased in recent years [30-36], and it is instructive to press forward in evaluating the merits of information theory and its role in animal movement modeling. Single, global estimations of correlation between animals are insufficient for modeling the complex relationships of animals over an extended time-period. Animals of the same or different species can have changes in the relationships of their movement as a result of mating behavior and critical resource allocation as well as seasonal and environmental changes in an ecosystem such as urbanization, deforestation, or climate change [37].

In this work, I propose a bandwidth-derived correlation function that accomplishing the following: (1) locally measures the correlation/association of the movement of two animals, (2) changes in correlation are successfully detected, (3) the measure can successfully detect complex relationships in a local domain, and (4) information about correlation in movement over a larger time-domain can be successfully inferred. The proposed correlation function is a measure of localized mutual information (LMI) using bivariate (longitudinal and latitudinal) animal position data. There is modest work in recent years in defining localized mutual information over a continuum, and the objective of this work is to construct a temporal measure of localized mutual information, thoroughly assess its qualities, and discuss the advantages and disadvantages of its implementation in movement ecology [38, 39].

I investigate the properties of the proposed LMI measure by proof, analytically, and by simulation. The measure has been previously implemented in a full statistical analysis of a collection of jaguars (Panthera onca) in the Pantanal Ecological Station in Brazil [40]. The measure is shown to be an effective tool for detecting shifts and spikes in animal association, and further, in the investigation of jaguar movement, the measure provides evidence of detecting animals in the same behavioral state, such as two male jaguars exhibiting similar movements in a mating season.

The simulation studies reveal several characteristics of the proposed measure of LMI which are of important consideration for researchers who intend to study associations in animal movement and behavior or use the measure for other telemetric, GPS tracking applications, such as military or citizen transportation movements, or relationships between animals and human transportation [41]. In the Discussion Section, following the simulations, I propose one alternative measure of LMI that addresses some of the disadvantages of the currently implemented measure.

## 2. Methods

### 2.1 Localized Mutual Information

Consider bivariate vectors $X = (X_1, X_2)$ and $Y = (Y_1, Y_2)$ which define the movement of two animals, where $X_1$ and $Y_1$ are vectors corresponding to the longitudinal position of the animals on a unified time grid, and $X_2$ and $Y_2$ correspond similarly for the latitudinal position. The proposed measure of LMI is defined by

$$\mathcal{I}(t;\lambda) = \mathbb{I}_{\mathcal{L}_i}(X, Y \mid \lambda) \quad \text{with} \quad \mathcal{L}_i = \{t \mid t \in [t_i - \lambda, t_i + \lambda]\}. \tag{1}$$

In brief, the LMI measure is constructed to compute the mutual information at every available time point $t_i$ for some local neighborhood defined by $t \in [t_i - \lambda, t_i + \lambda]$. The value of $\lambda$ is referred to throughout this document as the *bandwidth (bw)*. The bandwidth, chosen by the user, must be considered with knowledge of the application of interest. The output values of $\mathbb{I}_{\mathcal{L}_i}$ are a measure of joint mutual information for the directional components of their movements defined by

$$\mathbb{I}_{\mathcal{L}_i}(X, Y \mid \lambda) = \sqrt{I_{\mathcal{L}_i}(X_1, Y_1 \mid \lambda)^2 + I_{\mathcal{L}_i}(X_2, Y_2 \mid \lambda)^2}. \tag{2}$$

The measure of mutual information $I_{\mathscr{L}_i}(X_j, Y_j|\lambda)$ with $j = 1, 2$ is a measure of localized mutual information for the $j^{th}$ directional component defined by

$$I_{\mathscr{L}_i}(X_j, Y_j|\lambda) = \iint\limits_{\mathscr{X}_{j\mathscr{L}_i} \mathscr{Y}_{j\mathscr{L}_i}} p_{(X_j,Y_j)} \log \frac{p_{(X_j,Y_j)}(x,y)}{p_{X_j}(x)p_{Y_j}(y)} dxdy \quad \text{for } j = 1,2. \qquad (3)$$

The restriction of the mutual information to the localized neighborhood is given by $\mathscr{X}_{j\mathscr{L}_i}, \mathscr{Y}_{j\mathscr{L}_i}$. This is a restriction of the domain of $X, Y$ to their respective probability density functions on the time domain defined by $\mathscr{L}_i$. Since in most applications, there is expected longitudinal and latitudinal association between animals, the combinations of these associative components at each time point into a single metric is required in the construction of this definition.

## 2.2 General Properties

The LMI measure $\mathscr{I}(t; \lambda)$ is expected to carry many of the properties of a global measure of mutual information. Namely, in this section, I verify the non-negativity of the measure and the monotonicity of the relationship between LMI and a correlation coefficient $\rho$. Ultimately, it is of interest to ensure that any derived measure of association is on a scale that is comparable to a standard correlation coefficient, and further that the measure behaves similarly on such as scale.

*2.2.1 Proposition 1:* $\mathscr{I}(t; \lambda)$ is nonnegative (i.e. $\mathscr{I}(t; \lambda) \geq 0$).

P*roof:* It has been shown in previous literature that $I(\tilde{X}, \tilde{Y}) \geq 0$ where $I$ is a standard measure of mutual information and $\tilde{X}, \tilde{Y}$ are random variables [28]. For $\tilde{X}, \tilde{Y}$ defined with realizations on a unified time grid $\tau \in \mathbb{T}$ with $\tau = \{t_0, \ldots t_f\}$ and global time domain $\mathbb{T} \in \mathbb{R}$. For any subset $\{t_i\} \in \tau$ defined by $\mathscr{L}_i = \{t \mid t \in [t_i - \lambda, t_i + \lambda]\}$, $I_{\mathscr{L}_i}(X, Y) \geq 0$. Therefore for bivariate vectors $X = (X_1, X_2)$ and $Y = (Y_1, Y_2)$,

$$\mathbb{I}_{\mathscr{L}_i}(X, Y|\lambda) = \sqrt{I_{\mathscr{L}_i}(X_1, Y_1)^2 + I_{\mathscr{L}_i}(X_2, Y_2)^2} \geq 0.$$

Since $\mathscr{L}_i$ is arbitrarily defined, it follows that for any pair $(t_i, \mathscr{L}_i)$, $\mathscr{I}(t; \lambda) = \mathbb{I}_{\mathscr{L}_i}(X, Y|\lambda) \geq 0$. □

*2.2.2 Proposition 2:* As the association between X and Y increases (↑), $\mathscr{I}(t;\lambda)$ increases.

*Proof:* It is sufficient to show that the proposition holds for any $\mathscr{L}_i$ in the time domain $\mathbb{T}$. In other words, for any $\mathscr{L}_i$, it must be shown that as $\mathbb{I}_{\mathscr{L}_i}(X,Y|\lambda) \uparrow$, then the association between X and Y increases. It is known that the mutual information between two vectors X and Y on an interval $\mathscr{L}_i$ can be expressed by $I_{\mathscr{L}_i}(X,Y) = H_{\mathscr{L}_i}(X) + H_{\mathscr{L}_i}(Y) - H_{\mathscr{L}_i}(X,Y)$, where for any vector X, the Shannon's entropy, $H(X)$, is defined by $H(X) = -\int_{\mathscr{X}} p_X(x) log(p_X(x))\, dx$ [28]. Assume that $X \sim N(0,1)$ and $Y \sim N(0,1)$. Then it follows that

$$\begin{cases} H_{\mathscr{L}_i}(X) = \frac{1}{2} log\, 2\pi e \\ H_{\mathscr{L}_i}(Y) = \frac{1}{2} log\, 2\pi e \\ H_{\mathscr{L}_i}(X,Y) = \frac{1}{2} log((2\pi e)^2 (1-\rho^2)) \end{cases} \qquad (4)$$

where $\rho$ is the correlation coefficient between X and Y [28, 42]. It follows that

$$I_{\mathscr{L}_i}(X,Y) = \frac{1}{2} log\, 2\pi e + \frac{1}{2} log\, 2\pi e - \frac{1}{2} log((2\pi e)^2 (1-\rho^2)) \qquad (5)$$

$$I_{\mathscr{L}_i}(X,Y) = -\frac{1}{2} log(1-\rho^2). \qquad (6)$$

Refer to Figure 2.1 for a visual of the relationship Equation (6). It is clear that

$$\lim_{\rho \to 1} -\frac{1}{2} log(1-\rho^2) = +\infty, \quad \lim_{\rho \to -1} -\frac{1}{2} log(1-\rho^2) = +\infty, \text{ and}$$

$$\lim_{\rho \to 0} -\frac{1}{2} log((1-\rho)^2) = \lim_{\rho \to 0} -\frac{1}{2} log(1) = 0.$$

Since $I_{\mathscr{L}_i}$ is clearly monotonic in behavior on the domain $[-1,0) \cup (0,1]$, it follows that as $|\rho| \uparrow$, then $I_{\mathscr{L}_i} \uparrow$. If $X = (X_1, X_2)$ and $Y = (Y_1, Y_2)$, then.

$$\mathbb{I}_{\mathscr{L}_i}(X,Y|\lambda) = \sqrt{I_{\mathscr{L}_i}(X_1, Y_1) + I_{\mathscr{L}_i}(X_2, Y_2)} = \sqrt{-\frac{1}{2} log(1-\rho_1^2) - \frac{1}{2} log(1-\rho_2^2)}. \qquad (7)$$

Then if $|\rho_1| \uparrow$ and/or $|\rho_2| \uparrow$, then $\mathbb{I}_{\mathscr{L}_i}(X,Y|\lambda) \uparrow$. Since $\mathscr{L}_i$ is an arbitrary window within $\mathbb{T}$, then this holds for any $\mathscr{L}_i$ as defined by the pair $(t_i, \mathscr{L}_i)$. □

Although not examined in the proof of Proposition 2, it is important to emphasize that $X \sim N(0,1)$, $Y \sim N(0,1)$, and the integration of their respective pdfs assumes continuity when in the actual estimation pdf's empirically, this is computed by discretizing each pdf to a pmf.

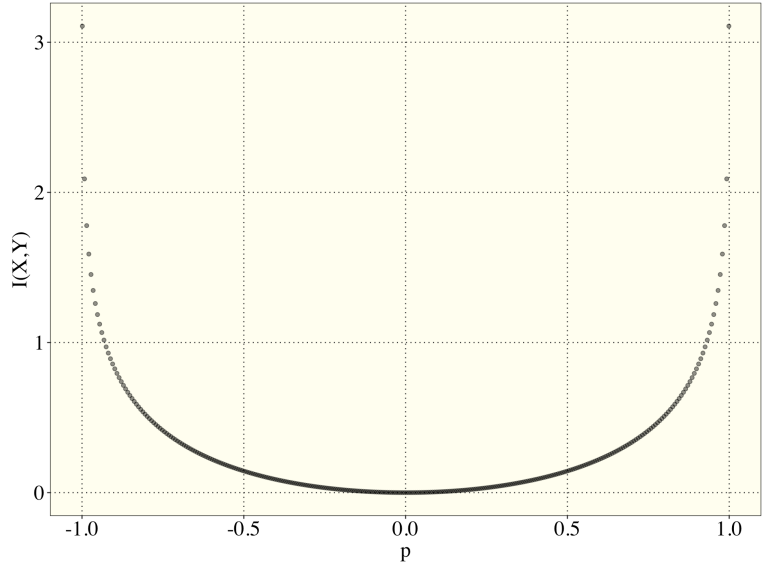

Figure 2.1: The relationship between mutual information and the correlation coefficient, $\rho$, for normally distributed $X, Y$ using a 300 equally spaced values of $\rho \in (-1, 1)$.

*2.2.3 Boundary Value Condition:*

The most apparent concern with the boundary cases for a bandwidth derived measure is the decreasing size of available samples of position within the bandwidth $\mathscr{L}_i$. If no boundary condition is coerced on $\mathscr{I}(t \mid \lambda)$, then at $t_0$, $\mathscr{L}_i = \{t \mid t \in [t_0, t_i + \lambda]\}$ and at $t_f$, $\mathscr{L}_i = \{t \mid t \in [t_i - \lambda, t_f]\}$ where $t_0$ and $t_f$ correspond to the first and last position of an animal detected or estimated. In this work, I set $\mathscr{L}_i$ for all *near boundary* $t_i$ according to the following condition. If $t_i - \lambda < t_0$, then $\mathscr{L}_i = \{t \mid t \in [t_0, t_0 + 2\lambda]\}$, and if $t_i + \lambda > t_f$, then $\mathscr{L}_i = \{t \mid t \in [t_f - 2\lambda, t_f]\}$. This boundary condition coerces every time point $t_i$ to have equal size $\mathscr{L}_i$ where any of the *near-boundary* $t_i$ have the same LMI. This is done to ensure that, as the bandwidth spans beyond the GPS operational period, the boundaries of the time domain are well-behaved and are not subject to fluctuations of LMI which result from the expected decline in observed positions.

## 3. Simulation Studies

In order to investigate the advantages and disadvantages of the proposed measure, it is instructive to generate simulations with known global mutual information and/or known associative structure over periods of constant behavior and report the LMI computed over these regions and those of transitioning behavior. More specifically, the sensitivity of the measure to the choice of bandwidth, $\lambda$, the number of bins (or simply *bins*) used to discretize a pmf for the movement paths over any $\mathscr{L}_i$, and the signal-to-noise ratio are the parameters of primary interest. The distance between the position of animals is of secondary interest, but is implicitly investigated here as well.

In the construction of simple movement simulations, I consider scenarios where LMI is both ideal and non-ideal, and a discussion of how to handle the non-ideal scenarios follows. Six simulation scenarios are investigated:

*Simulation 1:* A "Dead or Alive" movement model where the global mutual information, GMI, and LMI are feasible to derive by hand.

*Simulation 2:* A "Shift-Sensitivity" model where global mutual information is known for a period of strong association and a period of lower association.

*Simulation 3:* "Cross-directional Relationship" detection where two animals have negligent relationship between latitude's and longitude's but have a strong relationships between their opposite components.

*Simulation 4:* Delayed Onset/Following Behaviors where one animal follows the movement of the other.

*Simulation 5:* "Resolution Challenges" where the bandwidth and pdf discretization parameter sensitivity are examined in the presence of Brownian movement, and low versus high temporal resolution.

*Simulation 6:* A random walk simulation of male-female jaguar relationships [40].

In all simulations, the choice of bandwidth, $\lambda$, will be investigated, and in the other simulations, bins, and the signal-to-noise ratio will be examined at an array of values. The relevance of each of the movement behaviors is discussed in more detail in each of the following sections.

### 3.1 An analytically evaluated, simple simulation example

The preliminary simulation provides an analytic verification of the Global Mutual Information measure (GMI), the proposed LMI measure, and the agreement between the two measures on a simple animal movement simulation. The animal movement simulations progress in complexity, but in order to verify that the proposed measure is correctly detecting differences in association, I propose a simulation where two animals move linearly alongside each other for a period of time, and subsequently, one animal stops in its tracks for the second time window while the other animal continues to move linearly along its prior trajectory. The sudden halting movement behavior could characterize a resting, critically injured or deceased animal.

This simulation model for the first animal is defined by

$$\begin{cases} x(t_i) = t_i - 0.05 + \epsilon_{t_i} & t_i \in [0,1] \\ y(t_i) = t_i + \epsilon_{t_i} \end{cases},$$

and for the second animal,

$$\begin{cases} x(t_i) = t_i + 0.05 + \epsilon_{t_i} & t_i \in [0,0.5) \\ y(t_i) = t_i + \epsilon_{t_i} \\ x(t_i) = 0 + x(t_i = 0.5) + \epsilon_{t_i} & t_i \in (0.5,1.0] \\ y(t_i) = 0 + y(t_i = 0.5) + \epsilon_{t_i} \end{cases}.$$

In this basic simulation for which I derive an analytic solution, the random error term is set to 0. In most simulations, the error term is assumed to be independent and identically distributed. Intuitively, it is expected that if two animals move with the exact same behavior with no noise induced into the process, a perfect association in their movements will be detected. It also follows intuition that there should be no association in the movement of two animals if one animal is motionless and the other is moving with a clearly defined functional behavior.

A baseline analytic result for the mutual information of the two animal's movements is evaluated on each time interval as a global mutual information measure (GMI). In order to accomplish this, I examine the association in the longitudinal and the latitudinal movement in separate steps. Mutual information is directly related to the Shannon's entropy present in both random variables individually. More specifically, for two longitudinal random variables $X_1 = X_{lon1}$ and $X_2 = X_{lon2}$ the mutual information is defined as $I(X_2; X_1) = H(X_2) - H(X_2 | X_1)$, where $H(X_2) = - \sum_{X_2} P_{X_2}(x_2) log P_{X_2}(x_2)$ is the Shannon's entropy measure of the longitudinal movement of the first animal, and $H(X_2 | X_1) = \sum_{X_1} P_{X_1}(x_1)[-\sum_{X_2} P_{X_2|X_1}(x_2|x_1) log P_{X_2|X_1}(x_2|x_1)]$ is the conditional entropy of $X_2$ after observing the longitudinal movement in the other random variable $X_1$. Mutual information, by construction, is the reduction in uncertainty about a variable $X_2$ after $X_1$ is known.

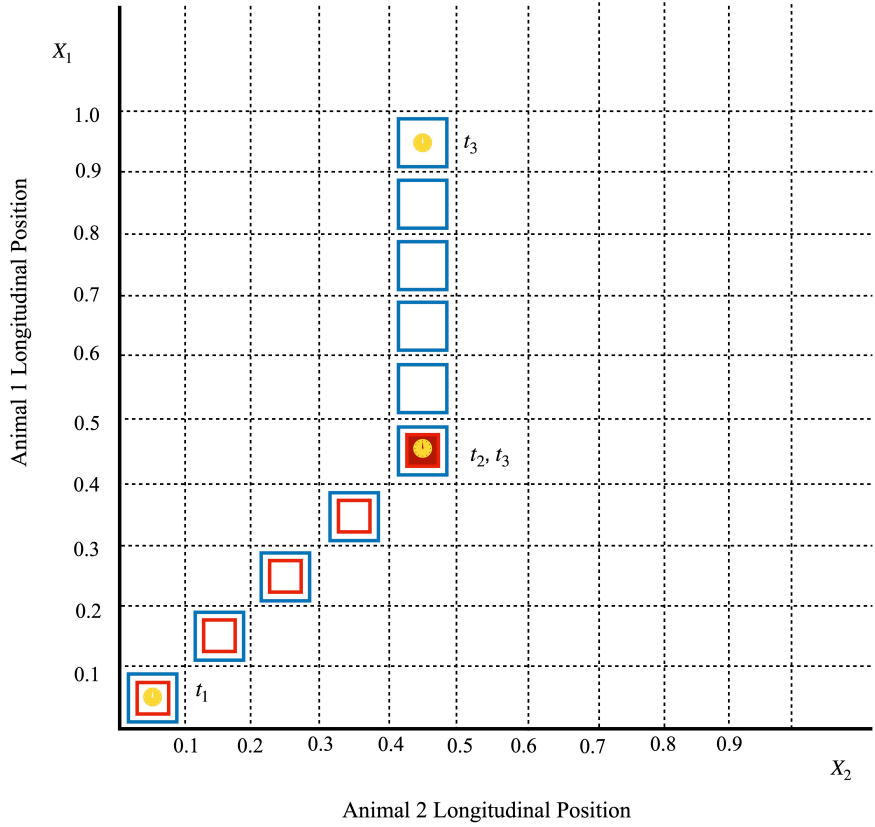

Figure 3.1: Longitudinal Movement of Both Animals discretized on a 0.1 resolution. Yellow circles denote the start and end time locations of the simulation as well as the time of the shift in behavior association.

The evaluation these sums is facilitated by discretizing the process defined for the longitudinal movement in the two models proposed above. Figure 3.1 provides an illustration of the discretization of this movement model. In Figure 3.1, blue and red squares identify the discretized location of animal 1 and animal 2 respectively, and the yellow circles marked the starting and end locations for each animal as well as the location where the shift in animal association occurs. The longitudinal behavior modeled here is identical to the constructed models, where animal 1's position is uniformly distributed within each blue square and animal 2's position is likewise, but all red squares from $t \in [t_2, t_3]$ are stacked in the same location. When computing the GMI for the time periods $[t_1, t_2]$ and $[t_2, t_3]$, the summation of the pmf's in the entropy components of the equation for mutual information consists of summing $N = 5$ squares (since 5 squares fall in each time period).

Beginning with $[t_2, t_3]$ for $X_1$ uniformly distributed with $P_{X_1}(X_1 = x_1) = \frac{1}{N}$. Then $X_2$ is singularly distributed since $X_2$ is within the same square for the entire interval; it is clear that $P_{X_2}(x = 0.5) = 1$. This is a consequence of animal 2 constant presence in this square for all 5 time-steps in this time interval. Since there is only one outcome in this time interval, it follows that

$$H(X_2) = -P_{X_2}(x_2 = 0.5)logP_{X_2}(x_2 = 0.5) = -1 \times log(1) = 0.$$

The conditional pmf of $X_2$ with $X_1$ observed is expressed by

$$P_{X_2|X_1} = \frac{P(X_2, X_1)}{P(X_1)} = \frac{1/N \times 1}{1/N} = 1, \text{ and}$$

$$H(X_2|X_1) = \sum_{n=1}^{N} P_{X_1}(x_1)[-\sum_{i=1}^{N} P_{X_2|X_1}logP_{X_2|X_1}(x_2|x_1)]$$

$$= \sum_{i=1}^{N} 1/N[-\sum_{i=1}^{N} 1 \times log(1)] = \sum_{i=1}^{N} \frac{1}{N}[0] = 0.$$

Then it follows $I(X_2; X_1) = H(X_2) - H(X_2|X_1) = 0 - 0 = 0$.

By construction, the latitude and longitude are characterized by the exact same behavior. For a pair of bivariate random variables $X_1 = (X_{Lon1}, X_{Lat1})$ and $X_2 = (X_{Lon2}, X_{Lat2})$, the LMI measure on a time window $\mathscr{L}_i = \{t \mid t \in [t_i - \lambda, t_i + \lambda]\}$ where $[t_i - \lambda, t_i + \lambda] \subset [t_2, t_3]$ is expressed by

$$\mathbb{I}_{\mathscr{L}_i} = \sqrt{I_{\mathscr{L}_i}(X_{Lon1}, X_{Lon2})^2 + I_{\mathscr{L}_i}(X_{Lat1}, X_{Lat2})^2} = \sqrt{0^2 + 0^2} = 0.$$

For the time $[t_1, t_2]$, both $X_1$ and $X_2$ are both uniformly distributed with $P_{X_1}(x_1) = P_{X_2}(x_2) = 1/N$. Then, the entropy of $X_2$ is expressed by

$$H(X_2) = -\sum_{i=1}^{N} P_{X_2}(x_2)log(P_{X_2}(x_2)) = -\sum_{i=1}^{N} 1/N \times log(1/N) = N \times 1/N\, log(N) = log(N).$$

The conditional entropy of $X_2$ with $X_1$ observed is

$$H(X_2|X_1) = \sum_{i=1}^{N} P_{X_1}(x_1)\left[-\sum_{i=1}^{N} P_{X_2|X_1}(x_2|x_1)\, log(P_{X_2|X_1}(x_2|x_1))\right]$$

$$= \sum_{i=1}^{N} P_{X_1}(x_1)\left[-\sum_{i=1}^{N} \frac{P_{X_2,X_1}(x_2,x_1)}{P_{x_1}(x_1)} log\left(\frac{P_{X_2,X_1}(x_2,x_1)}{P_{x_1}(x_1)}\right)\right]$$

$$= \sum_{i=1}^{N} 1/N \left[ -\sum_{i=1}^{N} \frac{1/N^2}{1/N} log\left(\frac{1/N^2}{1/N}\right) \right] = \sum_{i=1}^{N} 1/N \left[ \sum_{i=1}^{N} 1/N \, log(N) \right]$$

$$= \sum_{i=1}^{N} 1/N \times 1/N \times log(N) = 1/N \times log(N).$$

Then the mutual information is given by

$$I(X_2; X_1) = H(X_2) - H(X_2|X_1)$$

$$= log(N) - 1/N \times log(N) = log(N) \times (1 - 1/N)$$

and since $N = 5$, it follows that $I(X_2; X_1) = 0.8 \times log(5)$. Similar to before, with a pair of bivariate random variables $X_1 = (X_{Lon1}, X_{Lat1})$ and $X_2 = (X_{Lon2}, X_{Lat2})$, the LMI measure for any $t_i$ on an interval, defined by $\mathscr{L}_i$ such that $t_i - \lambda > t_2$, would be

$$\mathbb{I}_{\mathscr{L}_i} = \sqrt{I_{\mathscr{L}_i}(X_{Lon1}, X_{Lon2})^2 + I_{\mathscr{L}_i}(X_{Lat1}, X_{Lat2})^2} = \sqrt{(0.8 \times log(5))^2 + (0.8 \times log(5))^2}.$$

Furthermore, since $X_1$ and $X_2$ are both uniformly distributed, it follows that for any uniformly distributed variables defined on the same interval $[t_1, t_2]$, the mutual information would be the same answer. If we condition $X_2$ on itself, it is clear that this would defined perfect association of two pmfs, and it would also be the same magnitude of mutual information. Therefore the scaled mutual information, denoted by $\mathbb{I}^*$, by the maximum localized mutual information on this interval expressed by

$$\mathbb{I}^*(X_2; X_1) = \frac{\mathbb{I}(X_2; X_1)}{\mathbb{I}_{max}(X_2; X_2)} = 1.0.$$

Consequentially, the full simulation is characterized by a period of perfect association followed by a period of zero association as defined by our measure of mutual information. The LMI on any interval which overlaps the shift in behavior is bandwidth dependent. This is discussed in subsequent simulations, and it not explored analytically here. The scaling of LMI, $\mathbb{I}^*$, is used throughout this work to construct visualization of LMI on a scale of [0,1].

In this first simulation, the LMI is computed with several bandwidths. In the Figure 3.2, the plotted movement for both simulated animals is depicted. In Figure 3.3, the LMI is shown for six considered bandwidths on the arbitrary time domain of [0,1]. It is noted here that the largest bandwidth considered overlaps the behavior transition for almost all time points on the defined time grid. In these first simulations, random measurement error/noise is not induced into the process since the purpose of this simulation was to confirm agreeing LMI results from the analytic solution and the simulation.

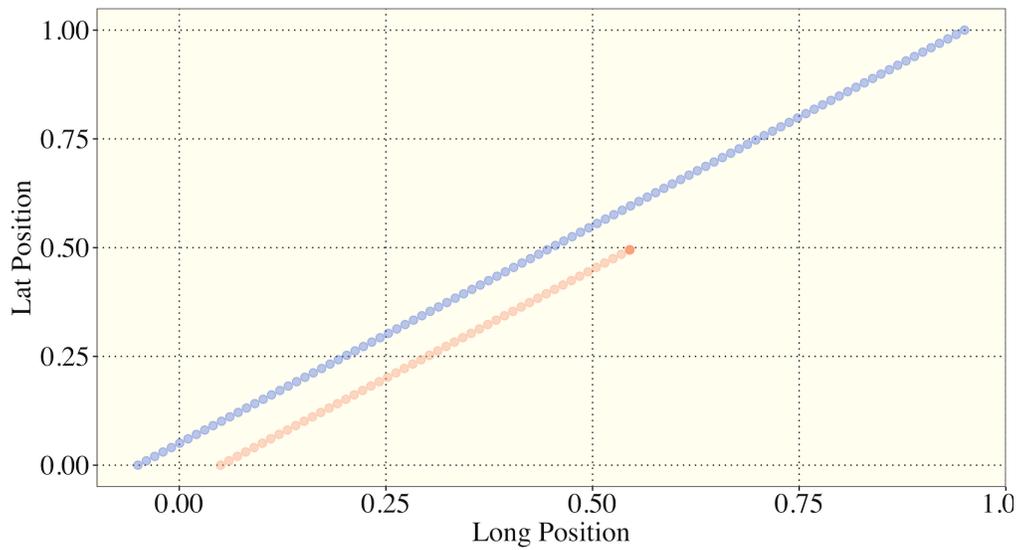

Figure 3.2: Simulated animals move in unison until the terminal halt of the second animal marked by a denser colored dot.

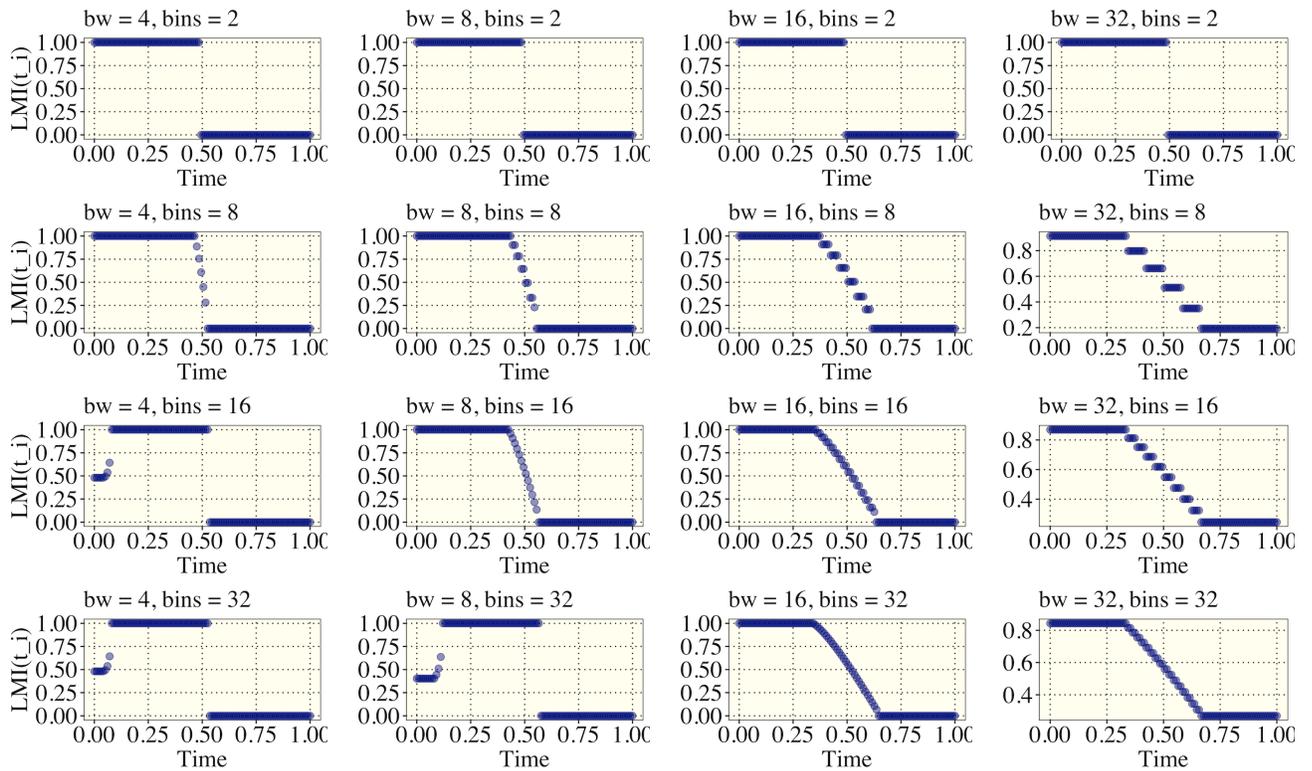

Figure 3.2. Note that in the 3 lower and left-most plots, poor end behavior conditions are a results of choosing a very small bandwidth with a larger number of bins used to discretize the animal pdfs. This is an expected outcome when computing mutual information with too few points relative to the number of bins.

In all cases the drop in association is clearly detected, and all of them contain valid information about the shift in association. Given the prior knowledge of the abruptness of the shift in behavior association, some researchers may prefer the upward and leftward plots of Figure 3.3 that show a steep drop in LMI to reflect the true nature of this correlatory shift. However, the gradual decline in LMI in the downward and rightward plots of Figure 3.3 which are using larger bandwidths indicates a constantly increasing number of time points that are in the zero-association region of the time domain. The "stair-steps" region of the plot marks time points that are centered on shifts between perfect and null association over portions of the time domain.

### 3.2 Simulation 2: Association Shift Sensitivity

Similar to the first simulation, where in the first half of the time domain both animals move linearly at the same speed, another simulation is constructed to assess the measure's ability to detect shifts in movement association that are similar functionally, but marked by a clear change in the relationship of two animals. The linear movement characterizing the animals in the first half of the time domains would be similar to herd/pack behavior or cooperative hunting in the case of solitary apex predators if the animals are close in proximity. If they are far apart from each other, this could characterize migratory movement induced by a similar climate dependent or seasonal shift in behavior.

The generation of independent and identically distributed noise with no propagated/auto-correlated error is important as a part of this simulation in order to ensure that the underlying movement generated is preserved, but at the same time partially confounded. Error defined in this way characterizes standard instrumentation error for GPS tracking devices where measurement error of some magnitude is inherent, and it is important that the LMI measure can still detect a shift in behavior in a simple movement scenario.

The generated movement of the first animal is defined by

$$\begin{cases} x(t_i) = t_i + a_{x1} + \epsilon_{t_i} & t_i \in [0,30] \\ y(t_i) = t_i + a_{y1} + \epsilon_{t_i} \end{cases},$$

and for the second animal

$$\begin{cases} x(t_i) = t_i + a_{x2} + \epsilon_{t_i} & t_i \in [0,15) \\ y(t_i) = t_i + a_{y2} + \epsilon_{t_i} \\ x(t_i) = b_{x2}(t_i - 15)^2 + x(t_i = 15) + \epsilon_{t_i} & t_i \in (15,30] \\ y(t_i) = -b_{y2}*(t_i - 15) + y(t_i = 15) + \epsilon_{t_i} \end{cases},$$

In our simulation, the coefficients are selected with an effort to generate movement where one animal returns to a the same region as the beginning of the simulation ($a_{x1} = -1$, $a_{y1} = 0$, $a_{x2} = 1$, $a_{y2} = 1$, $b_{x2} = -1/15$, $b_{y2} = -1.15$). The time domain is scaled to give an example when position is monitored hourly for a 30 day period, with a shift in behavior occurring at Day 15. The generated process characterizes perfectly associated movement from $t_i \in [0,15]$, as shown in the prior simulation, with a subsequent drop in movement association as the second animal turns sharply and moves with a different relationship between latitudinal and longitudinal movement. It has been confirmed that the scaled global mutual information from $t \in [0,15]$ is perfect/maximized ($GMI = 1.00$), and the global mutual information from $t \in (15,30]$ drops notably with the shift in behavior ($GMI = 0.93$).

The challenge of interest here is that the shift in movement association is stark, but the movement trajectories in the second time window are only subtly different. In brief, quadratic and linear functions are similar functionally, especially in small monotonic regions of the time domain for quadratic functions. As a result, the relatively high association between the two animals in the latter time window is expected since the strictly linear movement of Animal 1 can still sufficiently explain the quadratic movement of Animal 2. It is our objective in this first simulation to assess if the LMI measure proposed successfully detects this shift, and when it fails to effectively do so.

In Figure 3.4, the first round of simulations is performed with 4 different bandwidths, and zero error induced into the generated movement model. In this case, all bandwidths successfully detect the drop in movement association between the two animals, but smaller and larger bandwidths are detecting different but equally interesting features. The lower bandwidths of $bw = 2,4$, identify the largest drop in association on the [0,1] scale for LMI, but the larger bandwidth clearly marks the lowest point of association at the exact location of the shift. The reason for this is because time windows (centered on some $t_i$) that overlap the behavior shift are more disparate than time windows that don't overlap the transition. More specifically Animal 2's movement is not monotonic in the neighborhood surrounding the behavior shift, and LMI will record higher associations for regions where both animals have monotonic movement.

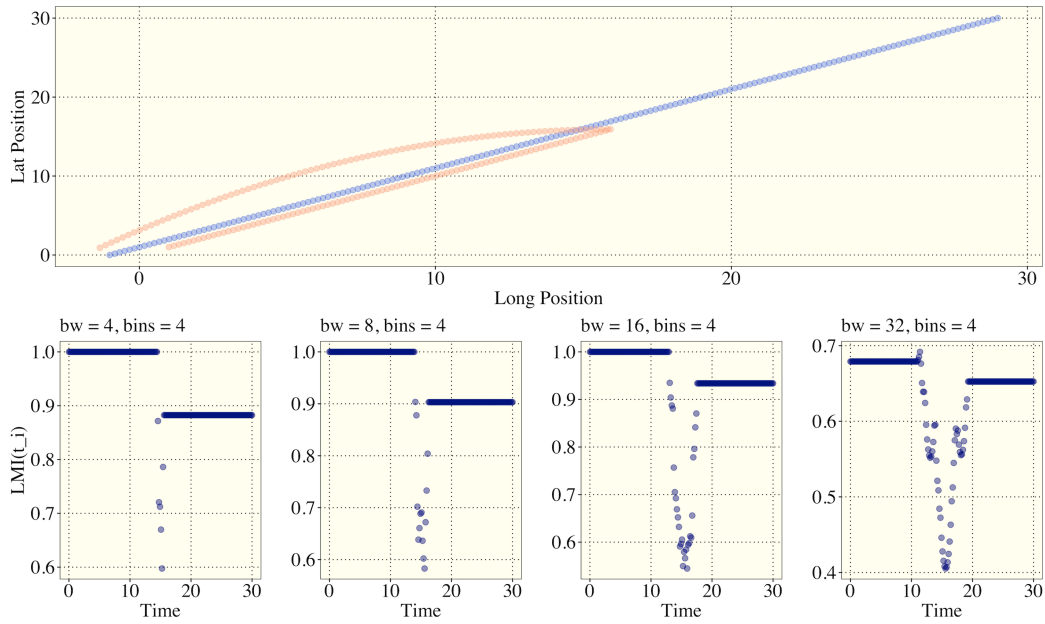

Figure 3.4: (Upper) Simulation 2 movement with $\epsilon_{t_i} = 0$. (Lower) LMI functions with various bandwidths (bw) and number of bins (bins) set to 4 for the discretized estimation of the pmf for each time window surrounding $t_i$.

In this movement scenario another round of simulations are performed in which the iid error is increased from $\epsilon \sim N(0,0)$ to $\epsilon\ N(0,1)$. In Figure 3.5, each set of simulations are shown with the same bandwidth parameters considers previously. All bandwidths satisfactorily detect a shift to lower association where the smaller bandwidths identify the shift as a stepwise/abrupt behavior shift. Bandwidth selection is studied in more detail in the Simulation 5 Section. It is apparent that substantial increases to noise don't inhibit the LMI measure's ability to detect shifts in behavior association, although the induced error prevents the model from detecting perfect association in the first half of the time domain. Three important conclusions from this simulation are summarized:

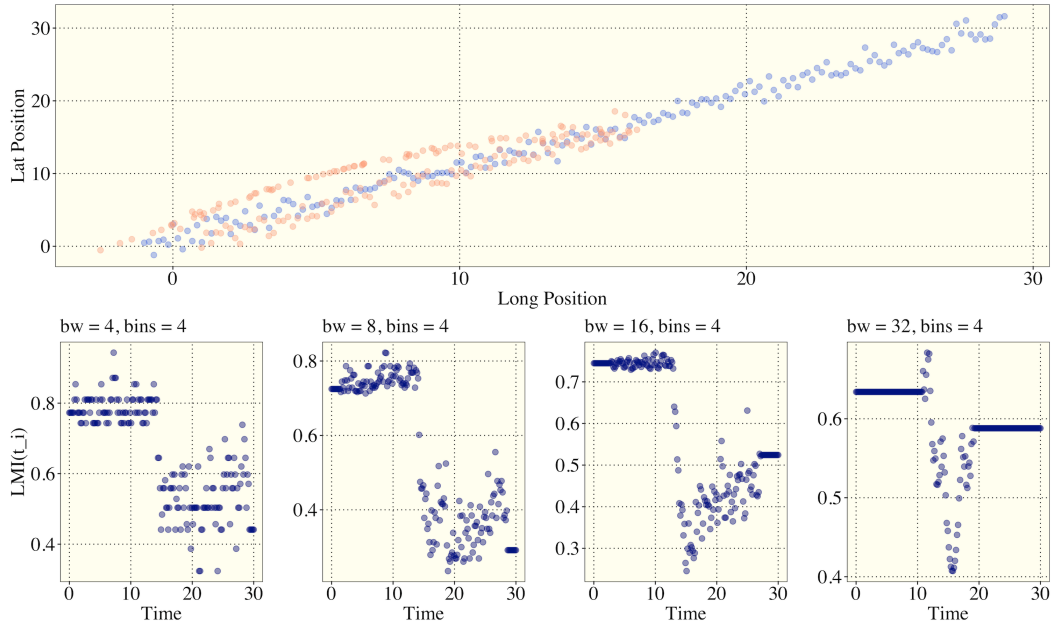

Figure 3.5: (Upper) Simulation 2 movement with $\epsilon_{t_i} = 0$. (Lower) LMI functions with various bandwidths (bw) and number of bins (bins) set to 4 for the discretized estimation of the pmf for each time window surrounding $t_i$.

1. Increasing random measurement error (which is assumed to be iid for each time point) marginally decreases the LMI measure's ability to detect an underlying process and association of animal movements. Any increase in the independence of a process will, by definition, lower the mutual information between two animals. If the errors have related/autocorrelated error terms then this issues would be clearly reduced. This is not shown since it is not the primary focus here, but the autocorrelation of error is possible if both animals are moving in a same region with poorer GPS signal. However, even in the presence of high iid error at each time, the measure still detects the shift and preserves the clear decline in association surrounding the time of the behavior shift.

2. Because of 1., it is important to note that it is advisable to smooth a process prior to using the proposed LMI measure. Smoothing splines, as an example, can be used to filter noise out of process that is known to be continuous, and the process is then represented by a Fourier or B-spline (polynomial) model which both have exceptional properties (such as differentiability) [43]. This was done in prior work with the LMI measure [40].

3. The proposed measure is highly sensitive to transition-states between behaviors. The LMI measure in all cases indicates that the time of lowest association between animals is in the direct neighborhood of the behavior shift as opposed to the entire last half of the time domain. This is attributed to the inability of linear movement to adequately contain information about an abrupt turning motion. As such, LMI shows evidence of being an exceptional measure for shift detection in the association of animal movements.

**Simulation 3: Cross-dimensional/directional Movement Association**

In this simulation, I highlight a nuance of LMI as defined in this work that can be addressed, when necessary, with a modification to the LMI measure. It has already been shown in Simulation 1 that any time $t_i$, if one animal is not moving while the other is moving, then we expect $\mathcal{I}(t = t_i; \lambda) = 0$. In the first simulation, this was true in the case where one animal was simulated to be stationary *in both* the latitude and longitude directions.

I propose another simple movement model where one animal moves due north, and the other animal moves due east. Animal 1 and Animal 2's movements are defined by

$$\begin{cases} x(t_i) = t_i + a_x + \epsilon_{t_i} & t_i \in [0,1] \\ y(t_i) = 0 + a_y + \epsilon_{t_i} \end{cases}, \text{ and } \begin{cases} x(t_i) = 0 + a_x + \epsilon_{t_i} & t_i \in [0,1] \\ y(t_i) = t_i + a_y + \epsilon_{t_i} \end{cases}.$$

Figure 3.6 present the resulting LMI from $t \in [0,1]$, and I only consider zero random error and some arbitrarily induced random error $\epsilon \sim N(0,0.05)$. Although perhaps a nuance, the results of this simulation indicate a clear weakness in the proposed measure of LMI. Since the joint (latitudinal and longitudinal) LMI measures only sums the association of a pair of animal movements by *matching directional components*, the measure is not detecting the strong *cross-directional* association. Ironically, in this extreme situation, inducing error into the process marks the animal movement behaviors as highly associative, although not perfectly associated as would be desirable.

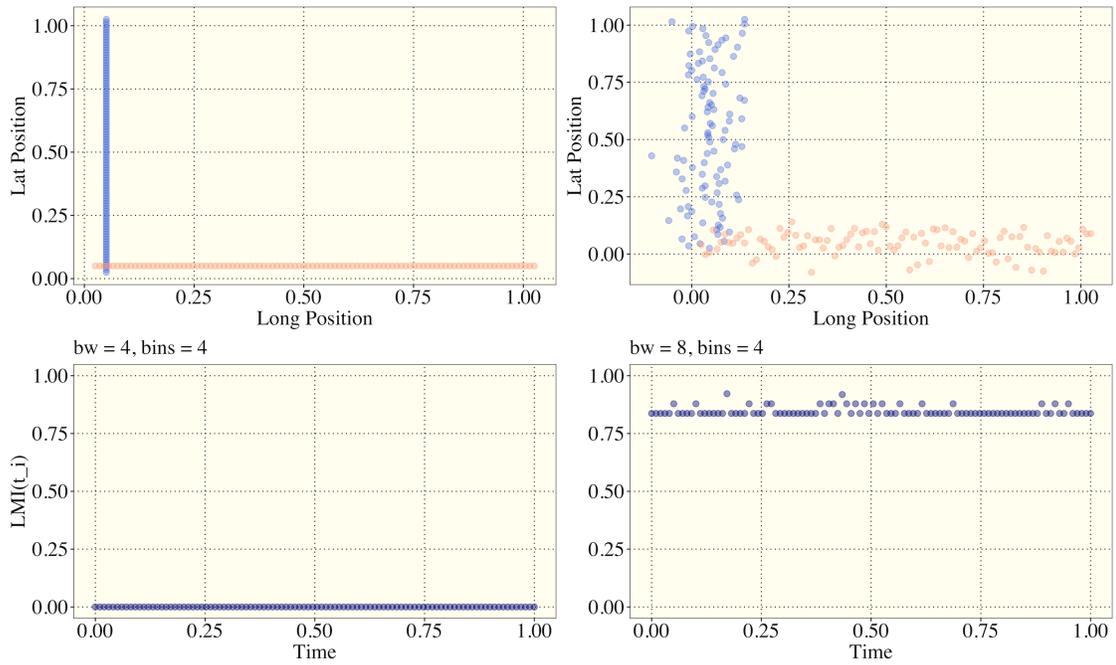

Figure 3.6: (Upper) Simulation 3 movement paths for strict cardinal movement generated with $\epsilon_{t_i} = 0$ and $0.05$ respectively. (Lower) The LMI Functions for the corresponding movement paths from the upper plots.

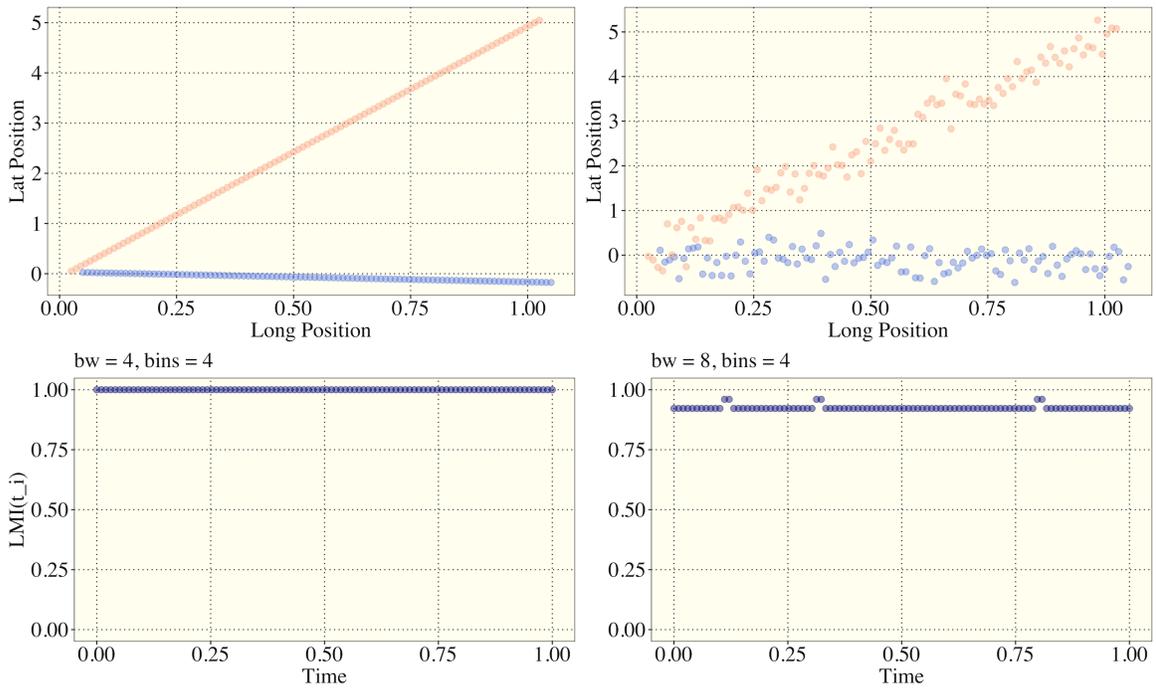

Figure 3.7: (Upper) Simulation 3 movement paths generated with $\epsilon_{t_i} = 0$ and $0.05$ respectively with rotation off the cardinal axes. (Lower) The LMI Functions for the corresponding movement paths from the upper plots.

Another similar movement model is proposed where the same movement is rotated off-parallel with the Longitude and Latitude axes. We now define Animal 1 and Animal 2's movements by

$$\begin{cases} x(t_i) = t_i + a_x + \epsilon_{t_i} & t_i \in [0,1] \\ y(t_i) = -0.20 * t_i + a_y + \epsilon_{t_i} \end{cases}, \text{ and } \begin{cases} x(t_i) = t_i + a_x + \epsilon_{t_i} & t_i \in [0,1] \\ y(t_i) = 5 * t_i + a_y + \epsilon_{t_i} \end{cases}.$$

These movement path's are perpendicular in $\mathbb{R}^2$ since both animals move at the same speed in the longitude-direction while Animal 1 and Animal 2's latitudes changes at a rate of -0.20 and +5 the rate of change in longitude. In Figure 3.7, the LMI measure is assessed with the same bandwidths and error terms. The rotation of movement yields a strong association which in comparison to the prior simulation marks an interpretation challenge for our measure of LMI that must be addressed. This is shown in Figure 3.7. For empirically observed or estimated movement trajectories, it is clear that witnessing a two animals movement exactly parallel to lines longitude and latitude is unrealistic, and so this measure is robust for most animal movement modeling problems.

However, this simulation still identifies an intriguing issue that stems from the definition of our current measure of LMI, which does not account for cross-directional movement association. I propose examples of where a better measure of LMI would be optimal:

1. Landscape structure at two distant sites: Consider two animals of the same species that reside along two different rivers. The first river runs approximately north-south and the second river runs approximately east-west. If the animals both move similarly along these rivers then the strongest relationship in their movement is expected to be cross-directional. This would be realistic for jaguars that hunt along similar waterways and may be located in disjoint habitats.

2. The example laid out in (1) would also hold for other land features. In Figure 3.8, an elevation map of the San Francisco Peaks of Northern Arizona is depicted. In this image, it is clear that the highest ridge is geographically curved, almost completing a full circle. If animals in this ecosystem have a tendency to move on the steep slopes of this range in the same way, then it is expected that two animals could have strong cross-directional association in their movements. Although, it would be highly unlikely to observe zero LMI (as in our simulation), if their movement associations are largely cross-directional, then a lower LMI would be reported than if they were on identically oriented terrain.

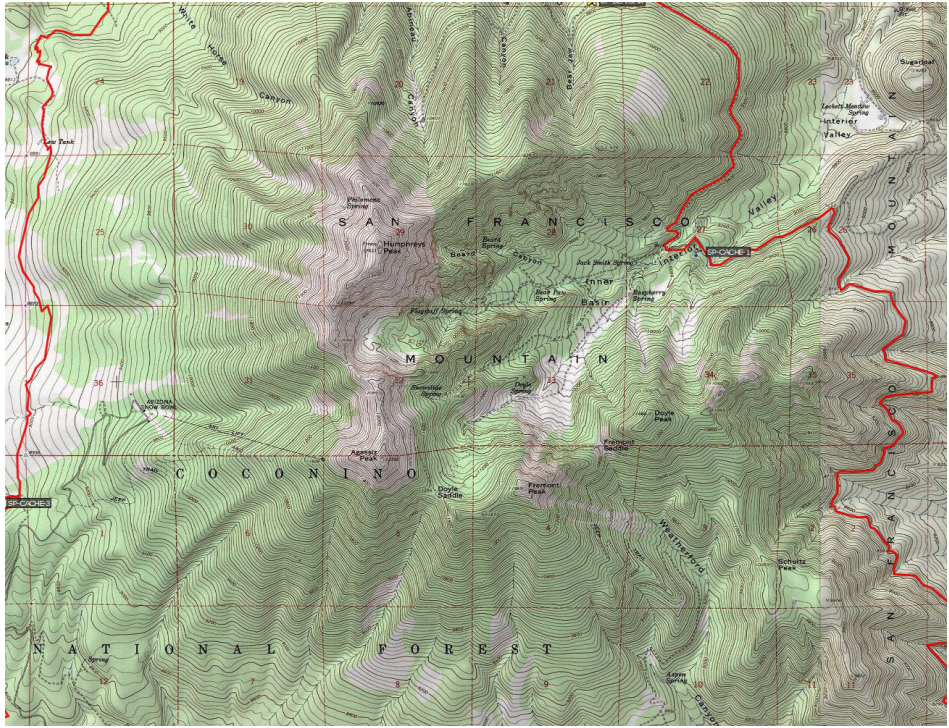

Figure 3.8: Topographic Map of the San Fransico Peaks and the surrounding section of the Coconino National Forest near Flagstaff, Arizona. The intent of presenting this figure is to draw attention to the natural C-shaped or bowl-shaped curvature of this mountain range.

3. Transportation data. Although not tied as closely to the field of movement ecology, the movement of planes, passenger vehicles, and trains etc. may be marked by cross-directionally related behavior, since the behavior of human transportation is largely more predictable. As an example, cars and trains will almost always follow roadways, railways, and air-routes, and many examples could be contrived to justify the presence of cross-directional association similar to (1) and (2). However, the integration of human and animal telemetry data has been considered in recent years which may mark an important future integrating human and animal movement and the detection of associations between such movements [40]. In the Discussion section, I propose a few simple modifications to our current measure of LMI to handle cross-directional association.

**Simulation 4: Delayed Onset/Tracking Movement Association**

In the first two simulations, animals that traveled alongside each other with the same behavior were shown to have perfect association as measured by the LMI measure. It becomes instructive to identify if the LMI measure identifies perfect association between animals that exhibit the same behavior but not at the same time. There are multiple scenarios where this would be important to consider in animal movement applications: (1) Predator prey relationships in many cases are characterized by following or tracking behavior by the predator. Examples include wolves following a herd of caribou or a mountain lion tracking one or multiple deer. In each of these cases, it would be of interest to researchers to identify how much time in a week, month, or season is marked by strong association between a subjects within predator and prey populations in an ecosystem. (2) Delayed regional migration of birds, such as arctic terns, could also be characterized in this manner. Recent literature has provided evidence that arctic tern colonies across the Northern Hemisphere share common migratory routes, and as such, the movements of some colonies along similar routes may be delayed by several days or weeks [47]. (3) Although not explored in this work, in the neotropical apex predator application that follows, jaguars, like many large cats, use scent and scrape marking strategies to passively communicate with each other [45, 46]. If predators are following similar routes to marked sites, then this could classified as a type of following/tracking behavior. The same may apply for males that are in search of females during a mating season.

In this simulation, the movement pattern complexity is increased to explore nonlinear movement. A simple example using "line drill/gyms suicides" motion is shown first. I model Animal 1's and 2's movements by

$$\begin{cases} x(t_i) = a_{xj} sin(t_i) + \epsilon_{t_i x} \\ y(t_i) = a_{yj} sin(t_i) + \epsilon_{t_y}, \end{cases} \text{and} \begin{cases} x(t_i) = a_{xj} sin(t_i - \varphi) + \epsilon_{t_i x} \\ y(t_i) = a_{yj} sin(t_i - \varphi) + \epsilon_{t_i y}. \end{cases}$$

For both animals, this movement is characterized by path retracing where additional time is spent at the ends of the path. In Figure 3.9, their movement is modeled to be identical except for a phase shift of $\varphi = \pi/2$. I consider two scenarios with $\epsilon = 0, 0.2$. Although clearly simplified, this movement characterizes a simple staggered migration between two locations, where an animal (such as a migratory bird) spends more time at the far ends of the route, and less time at any given position in between destinations.

In both cases, the LMI identifies perfect association in the animal movement paths with short periodic drops in their association, occurring at $t = \pi/2, \pi, 3\pi/2$. Over the domain of $t \in [0, 2\pi]$, these are the locations where the animals are briefly at points where animal 1 is either at an inflection point or a critical values on the sine function while the other animal is at the opposite. The magnitude of the drop in LMI computed during these short windows is dependent on the bandwidth, and the number of bins used in the discretization process although this is not shown in these simulations. The strong association is also well-captured in the presence of iid error.

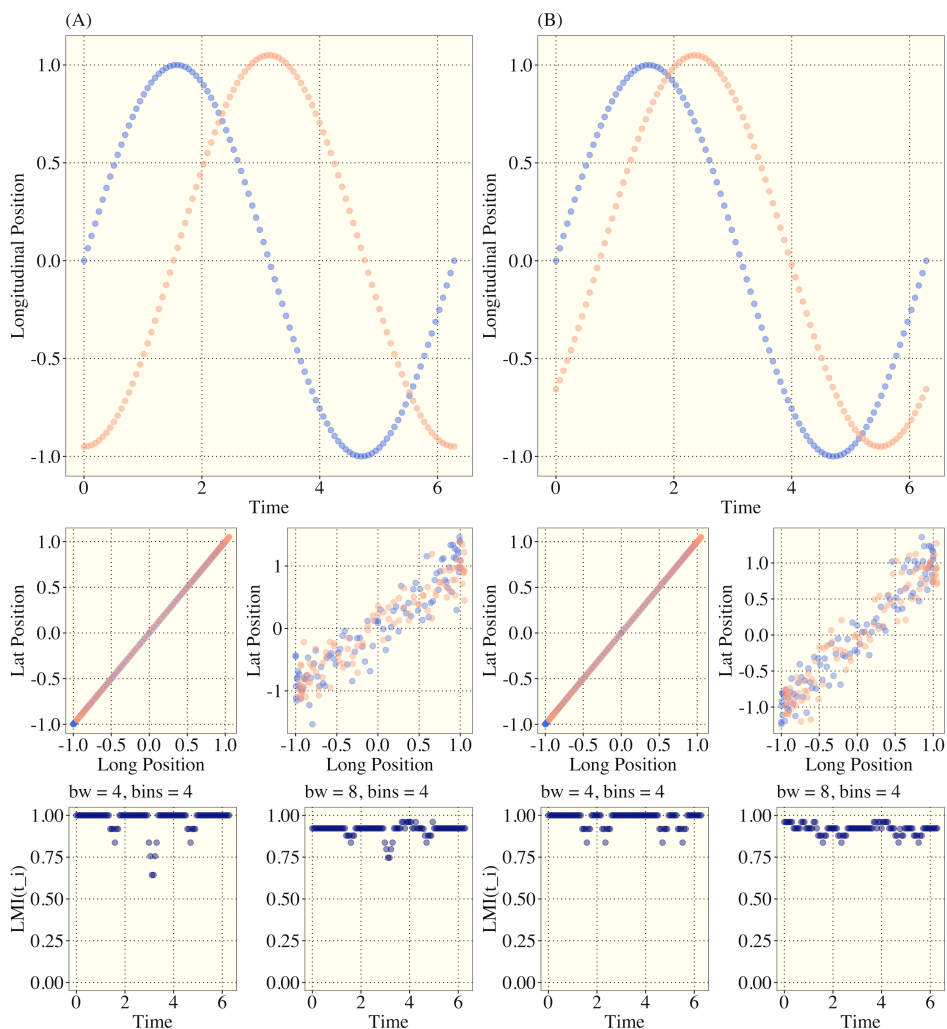

Figure 3.9: Simulation 4 Linear Tracing Movement Paths. (A) [Upper] plot show the induced temporal phase shift of $\varphi = \pi/2$ for the Longitudinal component. [Center] The generated movement paths in $\mathbb{R}^2$ with $\epsilon = 0, 0.2$. [Lower] The LMI functions for each of the respective errors. (B) Same as detailed for (A) but with $\varphi = \pi/4$.

In the second scenario, I consider the exact same movement behavior, but with a phase shift of $\varphi = \pi/4$. In the right column of plots in Figure 3.9, the periodicity of the drop from perfect association in LMI changes to $t = \pi/2, 3\pi/4, 3\pi/2, 7\pi/4$. At $t = \pi/2, 3\pi/2$, animal 1 is experiencing a direction change while the other animal continues to move in the same direction. At $3\pi/2, 7\pi/4$, animal 2 is experiencing the same (but delayed) direction change.

In both scenarios shown, the change in direction of either animal marks the dip in LMI. In essence, this is an identical phenomena to the drop in LMI shown at the behavioral shift in the Simulation 2 section; at an inflection point, the temporal neighborhood is monotonic, and at the critical values, the temporal neighborhood is not monotonic. More specifically, on a small window surrounding a direction change at a critical value, the surrounding time points characterized an even function (symmetric about $t_i$; $f(t) = -f(t)$). On a small time window at any other location, the surrounding time points characterize either an odd function ($f(t) = f(-t)$) in the case where the other animal is at an inflection point on the sine curve or the function is neither even nor odd.

Two more scenarios are considered for a more complicated moving pattern where two animals follow each other in a circle with some phase shift. Animal 1 and 2's movements are modeled by

$$\begin{cases} x(t_i) = a_{xj}sin(t_i) + \epsilon_{t_ix} \\ y(t_i) = a_{yj}cos(t_i) + \epsilon_{ty}, \end{cases} \text{ and } \begin{cases} x(t_i) = a_{xj}sin(t_i - \varphi) + \epsilon_{t_ix} \\ y(t_i) = a_{yj}cos(t_i - \varphi) + \epsilon_{t_iy}. \end{cases}$$

In these two scenarios, shown in the Figure 3.10, $\varphi = \pi/12$ and $\varphi = \pi/2$ respectively. The same conclusions from the prior scenarios are achieved here, where perfect association is detected except for at critical values in position for either animal, and in the presence of substantial iid random error, the same association is clearly detected. It is clear from the simulations in this section that proposed LMI measure is an exceptional tool for detecting following/tracking behavior. In order to eliminate the dips in LMI that would occur when either animal changes direction while the other continues along a monotonic trajectory, curve registration, warping, or time-lagging strategies would be required to align the curves. However, in many scenarios it is clear that this small dip may not occur biologically since tracing patterns may not resemble those shown in these simple simulations. Further, the instances where one animal changes direction while the other remains on a steady course may be important landmark features that should not be filtered out. As a result, this measure does not explicitly require curve alignment models to detect strong by delayed associative movements.

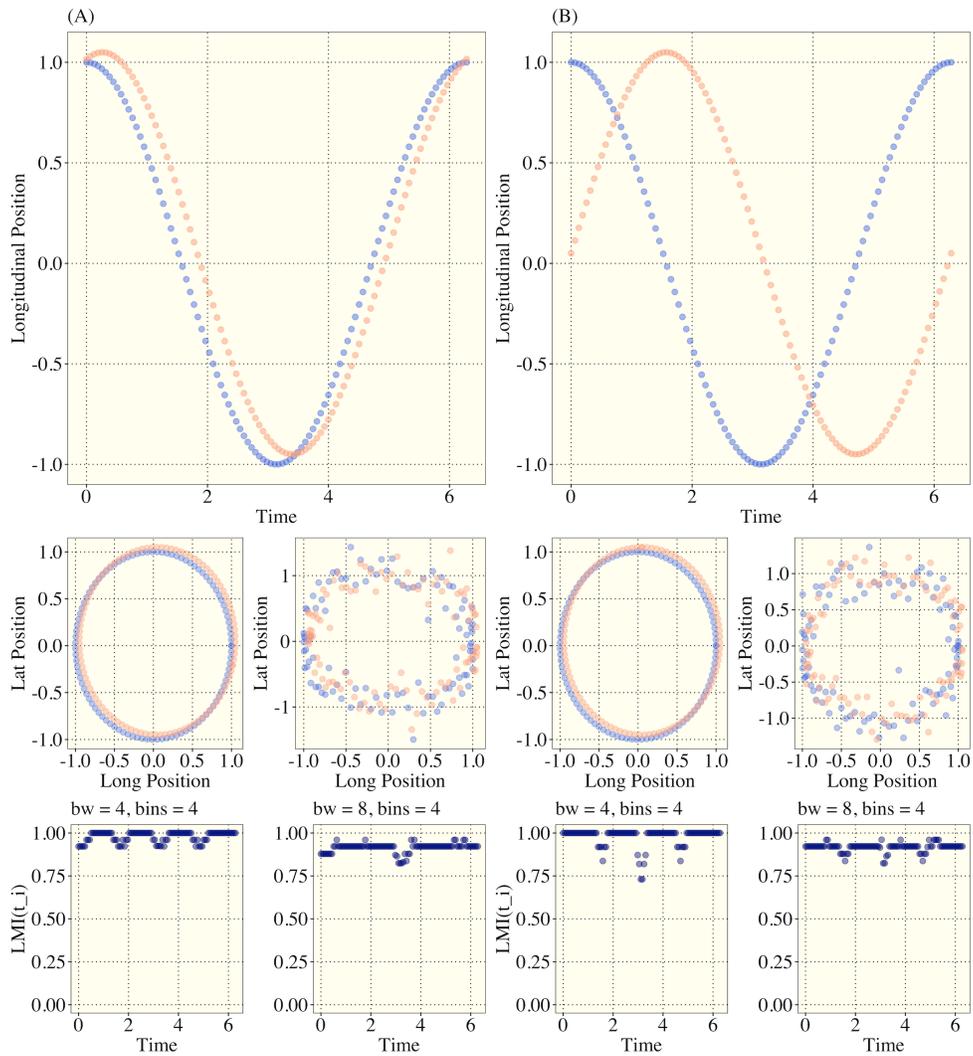

Figure 3.10: Simulation 4 Circular Tracing Movement Paths. (A) [Upper] plot show the induced temporal phase shift of $\varphi = \pi/2$ for the Longitudinal component. [Center] The generated movement paths in $\mathbb{R}^2$ with $\epsilon = 0, 0.2$. [Lower] The LMI functions for each of the respective errors. (B) Same as detailed for (A) but with $\varphi = \pi/4$.

**Simulation 5: Implications of Parameter Selection for Bandwidth and Bins**

In the Simulation 1 and Simulation 2 sections, there are clear (asymptotic) trends in the shape of the computed LMI function as the bandwidth and the number of bins used to construct the pmfs for each animal increases. There is also the question of what time resolution is required to uphold the integrity of the LMI measure. In the two sets of simulations in this section, I examine two distinct types of movement in a unit square. The first movement model simulates Brownian particle movement for both animals, and the second simulates identical but phase shifted cyclic movement for both animals similar to prior simulations in this work.

The Brownian movement models are both defined by

$$\begin{cases} x(t_i) = \epsilon_{t_i x} & t \in [0,1] \\ y(t_i) = \epsilon_{t_i y} \\ \epsilon_{t_i x}, \epsilon_{t_i y} \sim Unif(0,1). \end{cases}$$

For this movement model, characterized by jittered/random movement in the square $[0,1] \times [0,1]$, it is expected that no association should exist in the movements of two animals generate by this process. LMI is examined for time resolutions of 2, 4, 8, 16, 64, and 512 time points (including boundary times). I also consider bandwidths ranging from $bw \in [1,8]$ and number of bins ranging of 2, 3, and 4. The large collections of simulations described here are depicted in Figure 3.11.

For a small number of points, high and moderately high LMI is detected at most or all time points, and in the case of only having two points for each animal, the LMI measure always reports perfect association. As the time resolution (number of points) increases the LMI is gradually driven to zero. Further, increasing the number of bins increases LMI measure for all simulations where the time resolution permitted for an increase in bins. As an example, it would not make sense to discretize a movement path with only 4 observed locations using 8 bins, and the same concept applies for a local time window. The increases in LMI for increasing number of bins can be offset by increasing the bandwidth parameter. This is shown clearly in the final two rows of Figure 3.11.

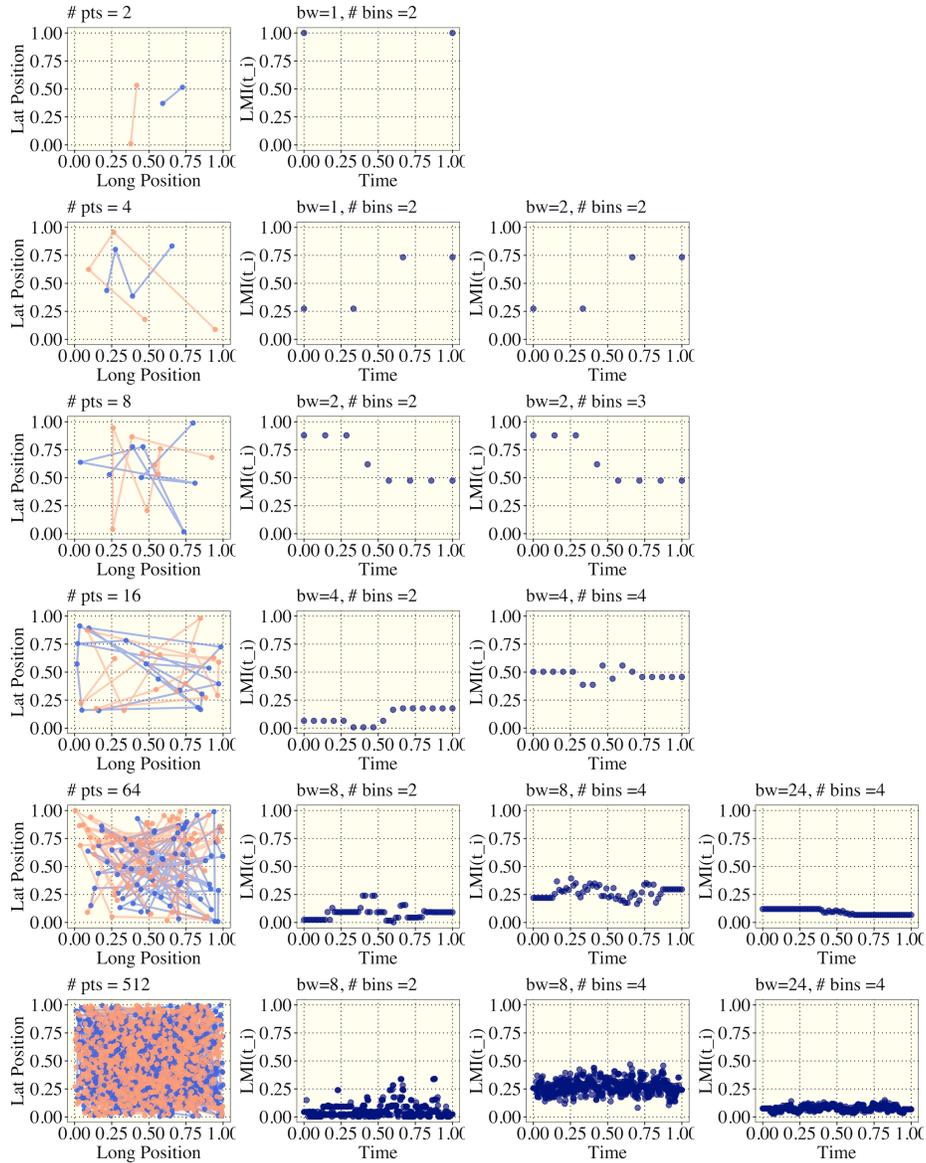

Figure 3.11: LMI Sensitivity to low temporal resolution for Brownian particle movement. The first column is the animal movement path in $\mathbb{R}^2$, and the remaining columns are the computed LMI functions for an array of bandwidths and bins.

The second set of simulations are intended to juxtapose the Brownian particle motion simulation. Animal 1 and 2 have movement defined by

$$\begin{cases} x(t_i) = a_{xj}sin(t_i) + \epsilon_{t_ix} \\ y(t_i) = a_{yj}sin(t_i) + \epsilon_{t_iy}, \end{cases} \text{ and } \begin{cases} x(t_i) = a_{xj}sin(t_i - \epsilon_\varphi) + \epsilon_{t_ix} \\ y(t_i) = a_{yj}sin(t_i - \epsilon_\varphi) + \epsilon_{t_iy}. \end{cases}$$

On the unit square, $\epsilon_{t_ix}, \epsilon_{t_iy} \sim N(0,0.025)$, and $\epsilon_\varphi \sim N(0,\pi/12)$. The random phase shifts, $\epsilon_\varphi$, allow for the animal to follow farther or closer that a fixed phase shift as shown in previous simulations. In Figure 3.12, the same parameters for bandwidth and number of bins are examined, and it is clear that the same problem exists for extremely low temporal resolutions where perfect or near perfect association is computed. However, as the temporal resolution increases the association between the animals stabilizes with clearer structure that is not trending towards zero like the Brownian motion simulation.

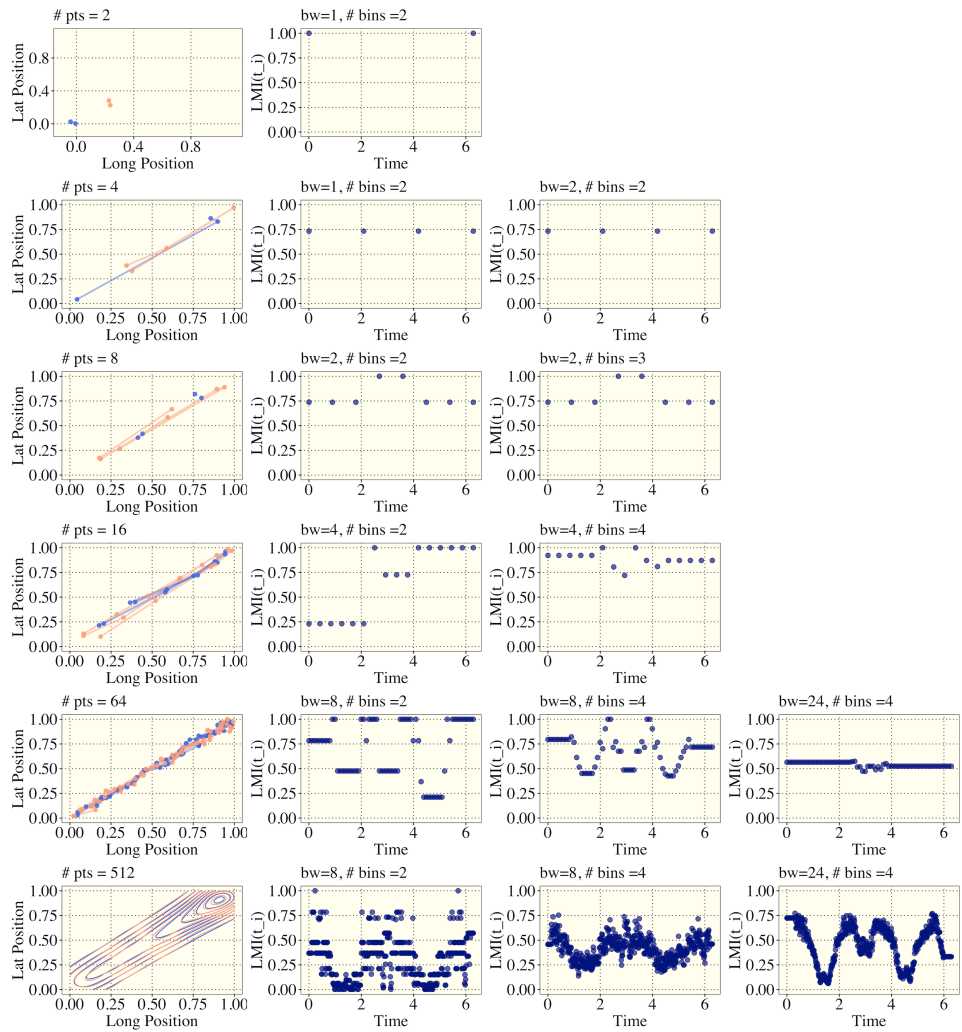

Figure 3.12: LMI Sensitivity to low temporal resolution for cyclical movement association. The first column is the animal movement path in $\mathbb{R}^2$, and the remaining columns are the computed LMI functions for an array of bandwidths and bins.

This section motivates guidelines for the use of this measure pertaining to required temporal resolution and tuning the bandwidth, and number of bins parameters. These guidelines are provided in the Discussion Section.

**Simulation 6: Simulating Male-female Neotropical Predator Movement Association**

The final set of simulations are motivated by the first implementation of this measure on monitoring male-female and male-male jaguar movement associations [40]. In one case detailed in this work, a male jaguar (Jaguar 18) relocated to the same region as a similar aged female (Jaguar 12). There were several periods of high cooccurence potential which were finally followed by Jaguar 12 removing herself by approximately 30km before returning to the same location two months later. At the end of the time window of high cooccurence potential, the proposed LMI measure detected a sharp decline to nearly zero using a bandwidth of 48 hours (or 4 days) [40]. I aim in this section to explore this specific behavior in a controlled simulation.

In the construction of an appropriate advanced animal movement model simulations, it is best to consider 3 primary objectives: (1) the simulated movement must mimic the complexity of animal movement in residential/courtship and migratory states, (2) the simulated movement should also be simplified to a single transition period to clearly illustrate shifts in the implemented LMI measure, and (3) the proximity of the male-female pairs must be reasonably captured in the residential/courtship and migratory state. For the residential/courtship state, the following random walk model is proposed for both male and female jaguars:

$$\begin{cases} x(t_i) = sin(t_i) + \epsilon_{t_i x} & t \in [0,30) \\ y(t_i) = cos(t_i) + \epsilon_{t_i y} \\ \epsilon_{t_i x} = \epsilon_{t_{i-1} x} + \omega_{t_i x} \\ \epsilon_{t_i y} = \epsilon_{t_{i-1} y} + \omega_{t_i y} \\ \omega_{t_i x} \text{ and } \omega_{t_i y} \underset{iid}{\sim} N(0,\sigma^2) \end{cases}$$

where $x(t_i)$ and $y(t_i)$ are the longitudinal and latitudinal position of the animal at time $i$, and $\epsilon$ is an autocorrelated error term associated with an animals movement at time $i$ and $i-1$.

This is a simplification of the solitary, but often related, non-stationary movements of male and female jaguars during courtship. This period of time is marked by frequent periods of high cooccurence potential with regular periods of separation. The modeled movement is circular by definition, but "drifting" movement is induced through the autocorrelated error terms $\epsilon_{t_i}$ which is an important consideration in modeling animal movement since deviations in foraging behavior may result in temporary or permanent shifts in movement trajectories. The initial condition/position of each jaguar is randomly instantiated within some arbitrary distance of the origin. As the movements of two jaguars under this simulated process will be marked by similar characteristics, the measure of LMI over this section of the time domain is expected to be high regardless of proximity.

In the following female migration state, the male jaguar will be set to continue his prescribed territorial movement as detailed previously. The female's movement behavior, however, will immediately transition to a migratory state modeled by

$$\begin{cases} x_f(t_i) = x_f(t_{i-1}) + \alpha \epsilon_{t_i x} + \beta |x_m(t_i)| + \xi_{t_i x} & t \in [30,60] \\ x_f(t_i) = y_f(t_{i-1}) + \alpha \epsilon_{t_i y} + \beta |y_m(t_i)| + \xi_{t_i y} \\ \epsilon_{t_i x} = \epsilon_{t_{i-1} x} + \omega_{t_i x} \\ \epsilon_{t_i y} = \epsilon_{t_{i-1} y} + \omega_{t_i y} \\ \omega_{t_i x} \text{ and } \omega_{t_i y} \underset{iid}{\sim} N(0,\sigma^2) \\ \xi_{t_i x} \text{ and } \xi_{t_i y} \underset{iid}{\sim} N(0,\sigma^2) \end{cases}$$

where $x_f$ and $y_f$ are the female's longitudinal and lateral position, and $x_m$ and $y_m$ are the male's longitudinal and lateral position which was generated from the first simulation model. This migratory model induces a repelling dependency between the male and female pair where the movement of the female is propelled based on the position of the male. The terms $\alpha, \beta = 0.01$ are scalars used to weight the value of the autocorrelated error term and the male position term to provide a realistic migration of the female that is not strictly linear since resource distribution and land cover may cause deviations from a direct straight-line migration away from her previous home range shared by the male. The initial condition for the female in this second time window, $t_0$, is set equal to the final position from the simulation generated for the first time window for the female to ensure that the generated path is a piecewise continuous process.

In Figure 3.13, the generated male and female movement models are shown in the 3 rightmost plots, and the autocorrelated error term progression is shown in the 2 leftmost plots. Both the male and female jaguars move in "drifting circles", and then the behavior transition of the female is stark and enduring for the entire remainder of the time domain. In the 60 day time period simulated, positions are generated every hour which yields 1440 points on the resulting time grid. In Figure 3.14, the LMI for their movement vectors is computed with $bw = 10, 50, 100$, and $bins = 2, 6, 10$. Recall that at any given time point $t_i$ the maximum attainable LMI is computed, and then the joint LMI measure is then scaled with respect to the maximum potential LMI, where $LMI(t_i) = 0.00$ identifies that no information about the movement of one jaguar is detectable from the other, and $LMI(t_i) = 1.00$ identifies that all information contained in one jaguar's movement is detectable from the other. The final LMI functions have 1440 points which are all depicted with a locally estimated scatterplot smoothing (LOESS) overlay to discern trends in cases where over-plotting of points in apparent.

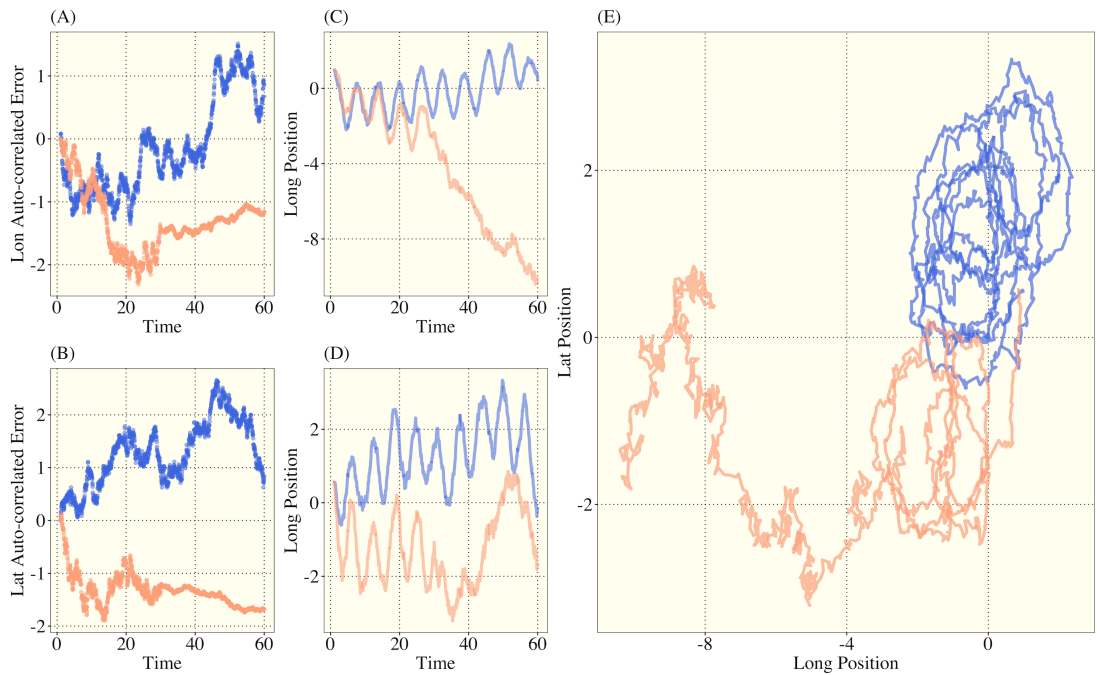

Figure 3.13: Simulated movement path for male (blue) and female (orange) jaguars with a behavior shift induced into the females movement at $t = 30$.

For all combinations of bandwidths and number of bins, the LMI measure successfully identifies strong association between the male-female pair during the time where both are in a residential state even though they are both subject to independently autocorrelated error terms. The LMI measure also successfully detects the shift in movement association. The detected shift in behavior is most apparent when increasing bandwidth and number bins, and the functional structure of the computed LMI functions is refined/de-noised. Of particular interest is the upper-right plot with $bw = 100$ and $bins = 2$. In this scenario, the drop in LMI is abrupt and bottoms out at exactly $t = 30$, where there is a "cusp-like" shift in behavior. With our prior knowledge of the abruptness of the induced shift in behavior, the LMI measure, with this choice of parameters, does an exceptional job at locating the sharp drop in movement association. As a reminder, $bw = 100$ indicates that the time window for each local measure of LMI is 200 hours, equivalently 8.3 days. All of the plots in the right most column shown declining association in 200 hour window LMI measures. There is a peculiar spike in LMI shown in these plots which can be attributed to the similar latitudinal behavior of both animals for approximately $t \in [40,50]$. This can be seen in the bottom-center plot of Figure 3.13, where both the male and female have repeated instances where their latitudinal position is increasing in unison.

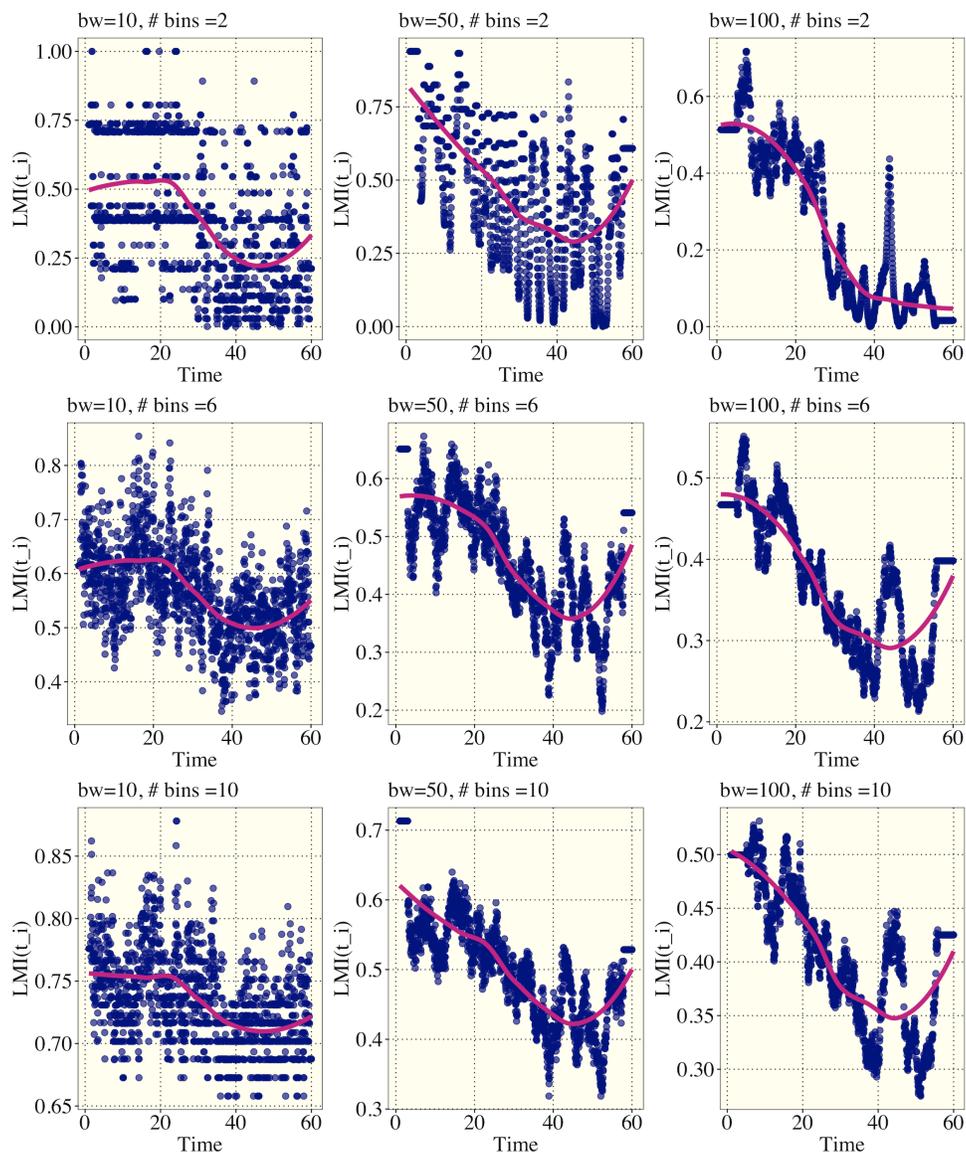

Figure 3.14: LMI function computed for an array of bandwidths and number of bins. The overlaid magenta line in each plot is a LOESS trend line.

## 4. Discussion

This work details a rigorous process of evaluating a new measure of association between GPS detected movement paths or estimated trajectories. I summarize the important details of this work with commentary when necessary.

1. Non-negativity: the LMI measure is shown by proof to be semi-positive definite.

2. Monotonic relationship with $\rho$. Although shown for only normally distributed random variables, it is shown that increases in correlation between two variables also indicates an increase in the LMI measure. This would need to be shown for other distributions more generally in future work to confirm that this is true for a wider range of distributions from the exponential family.

3. Shift Detection: The LMI measure is an effective tool for detecting a shift in the association of animal movements even if the shift is functionally subtle. Shifts from monotonic to non-monotonic movement for only one of the animals yield large drops in movement association in the neighborhood surrounding such behavior.

4. Cross-directional association: Perhaps the primary disadvantage of the presently evaluated measure of LMI is the absence of cross-directional association detection. I propose a simple modification to the current measure that would address this disadvantage. The new measure is expressed by

$$\begin{cases} \mathcal{I}(t;\lambda) = \mathbb{I}_{\mathcal{L}_i}(X,Y \mid \lambda) \text{ with } \mathcal{L}_i = \{t \mid t \in [t_i - \lambda, t_i + \lambda]\}, \\ \text{where } \mathbb{I}_{\mathcal{L}_i}(X,Y) = \nu_i \times (I_{\mathcal{L}_i}(X_1, Y_1) + I_{\mathcal{L}_i}(X_2, Y_2)) + (1 - \nu_i)(I_{\mathcal{L}_i}(X_1, Y_2) + I_{\mathcal{L}_i}(X_2, Y_1)), \\ \nu_i \in [0,1] \\ I(X_j, Y_j) = \iint_{\mathcal{X} \times \mathcal{Y}} p_{(X,Y)} \log \frac{P(X,Y)}{p_X(x) p_Y(y)} dx dy \text{ for } j = 1,2. \end{cases}$$

The terms in this model are not squared or square-rooted as in the measure of LMI defined in this project, which is not necessarily needed since the mutual information components used to compute LMI are shown to be non-negative. The critical modifications in this proposed measure are the incorporation of two additional cross-directional measures of mutual information for the relationship between animal 1 and animal 2's longitude and latitude respectively and vice versa. An important question of the value of cross-directional versus standard mutual information on a time window $\mathcal{L}_i$ is more important to detect and visual in the final outputted LMI function visualizations. The parameter $\nu$ could be defined uniformly across the full time domain to value cross-directional and standard LMI equally with $\nu = 0.5$ or with $\nu = 0,1$ the standard or cross-directional LMI would be completely disregarded. The $\nu_i$, corresponding to a specific time window $\mathcal{L}_i$, could be specifically set for certain portions of the time domain to up-weight cross-directional mutual information when (as an example) the aspect of the slope that two animals are on are estimated to be near perpendicular.

4. Delayed Onset tracking movement: The LMI measure is an appropriate tool for detecting following or path-tracing/tracking behavior. This attribute of the measure will have practical uses in monitoring larger volumes of animal migration data and predator-prey dynamics, as well as human-animal interaction applications in cases, where human movement/transportation can be tracked over the same time period.

5. Tuning Parameters: Giving the user parameters to subjectively tune could be considered a disadvantage, but I argue against this notion as it is an opportunity for the researchers, implementing such a measure, to use field expertise to adjust the measure to the application or behavior of interest. The parameters are not abstract: *bandwidth* controls the size of the temporal neighborhood used to measure correlation between animals at a given time, and *number of bins* controls the complexity of discretized movement over that same neighborhood. As mention in (4), locally adaptive parameters would be an interesting extensions of this measure.

As a whole, this work justifies the use of bandwidth derived correlation functions. Mutual information is a highly flexible measure of association that can handle violations of monotonicity that challenge the use of Pearson and Spearman correlation measures. I return to our ultimate proposed question: *What does correlation mean in animal movement?*. Correlation is ultimately a measure of dissimilarity of two processes or observational units. In movement ecology, any characteristics that make the movements of animals more similar should be detectable by some dissimilarity metric. Correlation in movement paths or trajectories should be a flexible idea, and there should be several types of movement behaviors that are identifiable prior to a full statistical modeling process. In any standard regression analysis it is common practice to examine a correlation matrix of predictor and response variables and the variance inflation factor. We should model this process in movement ecology by considering metrics that detect association between movement trajectories of all animals in the study. I emphasize that this notion of dissimilarity is a means as opposed to an end objective. Dissimilarity measures are a vital tool in modern statistical modeling, especially unsupervised machine learning and spatial statistics.

I encourage the prescribed use of this measure, and more importantly, I encourage others to challenge the understanding of movement association in telemetric data analysis laid out in this work.

**Acknowledgements**

The quality of this work has been aided greatly by the efforts of multiple individuals: Ronaldo Morato for his ongoing communications during my original implementation of the measure of jaguar pairs in the Pantanal Ecological Station, Daniel Gervini for his early contributions and insightful criticism of this work, and Vincent Larson for his invaluable encouragement, theoretical contributions and numerous discussions of methods results and implications.

**References**

1. Urbano, F., Cagnacci, F., Calenge, C., Dettki, H., Cameron, A., Neteler, M. (2010). Wildlife tracking data management: a new vision. Philosophical transactions of the Royal Society of London. Series B, Biological sciences, 365(1550), 2177–2185. https://doi.org/10.1098/rstb.2010.0081
2. Lewis M.A., Fagan W.F., Auger-Méthé M., Frair J., Fryxell J.M., Gros C., Gurarie E., Healy S.D., Merkle J.A. Learning and Animal Movement. Frontiers in Ecology and Evolution. Vol 9. pp 441. https://www.frontiersin.org/article/10.3389/fevo.2021.681704.
3. Cagnacci F., Urbano F.2008Managing wildlife: a spatial information system for GPS collars data. Environ. Modell. Softw. 23, 957–959 (doi:10.1016/j.envsoft.2008.01.003)
4. Hooten, M., Buderman, F., Brost, B., Hanks, E., and Ivan, J. (2016), "Hierarchical Animal Movement Models for Population-Level Inference," Environmetrics, 27, 322–333.
5. Hooten M.B., Johnson D.S., McClintock B.T., Morales J.M., References, Animal Movement, 10.1201/9781315117744, (273-290), (2017).
6. Fleming, C., Deznabi, I., Alavi, S., Crofoot, M., Hirsch, B., Medici, E., Noonan, M., Kays, R., Fagan, W., Sheldon, D., Calabrese, J. (2021). Population-level inference for home-range areas. 10.1101/2021.07.05.451204.
7. Buderman F.E. et. al. A functional model for characterizing long-distance movement behaviour. Methods in Ecology and Evolution. Volume7, Issue3 March 2016. Pages 264-273. https://doi.org/10.1111/2041-210X.12465
8. McClintock, B.T., Johnson, D.S., Hooten, M.B. et al. When to be discrete: the importance of time formulation in understanding animal movement. Mov Ecol 2, 21 (2014). https://doi.org/10.1186/s40462-014-0021-6
9. McClintock, B., London J., Cameron M., and Boveng P. 2015. Modeling animal move- ment using the Argos satellite telemetry location error ellipse. Methods in Ecology and Evolution, 6:266–277.
10. Jonsen, I., Flemming J., and Myers R. 2005. Robust state-space modeling of animal movement data. Ecology, 45:589–598.
11. Morales, J., Haydon D., Friar J., Holsinger K., and Fryxell J. 2004. Extracting more out of relocation data: building movement models as mixtures of random walks. Ecology, 85: 2436–2445.
12. Parton, A., Blackwell P.G. Bayesian Inference for Multistate 'Step and Turn' Animal Movement in Continuous Time. JABES 22, 373–392 (2017). https://doi.org/10.1007/s13253-017-0286-5
13. Hanks, E.M., Hooten M.B., and Alldredge M.W. "Continuous-time discrete-space models for animal movement." The Annals of Applied Statistics 9.1 (2015): 145-165
14. Harris KJ, Blackwell PG. Flexible continuous-time modelling for heterogeneous animal movement. Ecol Model 2013, 255:29–37
15. Johnson, D.S., et al. "Continuous‐time correlated random walk model for animal telemetry data." Ecology 89.5 (2008): 1208-1215
16. Gurarie E, Andrews RD, Laidre KL. A novel method for identifying behavioural changes in animal movement data. Ecol Lett. 2009; 12: 395–408. pmid:19379134
17. Torres LG, Orben RA, Tolkova I, Thompson DR (2017) Classification of Animal Movement Behavior through Residence in Space and Time. PLOS ONE 12(1): e0168513. https://doi.org/10.1371/journal.pone.0168513

18. Sur M, Skidmore AK, Exo K-M, Wang T, Ens BJ, Toxopeus A. Change detection in animal movement using discrete wavelet analysis. Ecol Inform. 2014; 20: 47–57.
19. Fontes S.G., Morato R.G., Stanzani S.L., Pizzigatti Corrêa P.L. (2021) Jaguar movement behavior: using trajectories and association rule mining algorithms to unveil behavioral states and social interactions. PLOS ONE 16(2): e0246233. https://doi.org/10.1371/journal.pone.0246233
20. Hastie, T., Tibshirani, R.,, Friedman, J. (2001). The Elements of Statistical Learning. New York, NY, USA: Springer New York Inc..
21. Khaled Fawagreh, Mohamed Medhat Gaber, Eyad Elyan (2014) Random forests: from early developments to recent advancements, Systems Science and Control Engineering, 2:1, 602-609, DOI: 10.1080/21642583.2014.956265
22. Gower, J.C., Properties of Euclidean and non-Euclidean distance matrices. Linear Algebra and its Applications. Volume 67. 1985. Pages 81-97. ISSN 0024-3795,https://doi.org/10.1016/0024-3795(85)90187-9.
23. Boriah S., Chandola V., Kumar V. (2008). Similarity measures for categorical data: A comparative evaluation. In: Proceedings of the 8th SIAM International Conference on Data Mining, SIAM, p. 243-254.
24. Whetten AB, Demler H. Detection of Multidecadal Changes in Vegetation Dynamics and Association with Intra-annual Climate Variability in the Columbia River Basin. In ArXiv e-prints (May 2021). arXiv: 2105.08864 [q-bio.QM]
25. Binbin Lu, Martin Charlton, Paul Harris, A. Stewart Fotheringham (2014) Geographically weighted regression with a non-Euclidean distance metric: a case study using hedonic house price data, International Journal of Geographical Information Science, 28:4, 660-681, DOI: 10.1080/13658816.2013.865739
26. Kendall, M. G. (1938). A new measure of rank correlation, Biometrika, 30, 81–93. doi: 10.1093/biomet/30.1-2.81.
27. Schaeffer, M. S.,  Levitt, E. E. (1956). Concerning Kendall's tau, a nonparametric correlation coefficient. Psychological Bulletin, 53(4), 338–346. https://doi.org/10.1037/h0045013
28. Cover, T.M.; Thomas, J.A. (1991). Elements of Information Theory (Wiley ed.). ISBN 988-0-471-24195-9.
29. Gervini, D. and Khanal, M. (2019). Exploring patterns of demand in bike sharing systems via replicated point process models.  Journal of the Royal Statistical Society Series C: Applied Statistics 68 585-602.
30. Owoeye K., Musolesi M., Hailes S. Characterizing animal movement patterns across different scales and habitats using information theory. bioRxiv 311241. 2018. doi: https://doi.org/10.1101/311241
31. S. Butail, F. Ladu, D. Spinello, and M. Porfiri. Information flow in animal-robot interactions. Entropy, 16(3):1315–1330, 2014.
32. S. Butail, V. Mwaffo, and M. Porfiri. Model-free information-theoretic approach to infer leadership in pairs of zebrafish. Physical Review E, 93(4):042411, 2016
33. F. Hu, L.-J. Nie, and S.-J. Fu. Information dynamics in the interaction between a prey and a predator fish. Entropy, 17(10):7230–7241, 2015.
34. M. Kadota, E. J. White, S. Torisawa, K. Komeyama, and T. Takagi. Employing relative entropy techniques for assessing modifications in animal behavior. PLOS ONE, 6(12):1–6, 2011.
35. W. M. Lord, J. Sun, N. T. Ouellette, and E. M. Bolt. Inference of causal information flow in collective animal behavior. IEEE Transactions on Molecular, Biological and Multi-Scale Communications, 2(1):107–116, 2016.
36. Rocchini, D.; Thouverai, E.; Marcantonio, M.; Iannacito, M.; Da Re, D.; Torresani, M.; Bacaro, G.; Bazzichetto, M.; Bernardi, A.; Foody, G.M.; et al. rasterdiv—An Information Theory tailored R package for measuring ecosystem heterogeneity from space: To the origin and back. Methods Ecol. Evol. 2021, 12, 1093–1102.
37. Jeffrey J. Thompson, Ronaldo G. Morato, et. al. (2021) Environmental and anthropogenic factors synergistically affect space use of jaguars. Current Biology. 31 (15): 3457-3466. ISSN 0960-9822. https://doi.org/10.1016/j.cub.2021.06.029.
38. Dai S., Zhan S., Song N. Adaptive Active Contour Model: a Localized Mutual Information Approach for Medical Image Segmentation. (2015). KSII Transactions on Internet and Information Systems, 9(5). https://doi.org/10.3837/tiis.2015.05.016
39.  Klein S., et. al. Automatic segmentation of the prostate in 3D MR images by atlas matching using localized mutual information.2008. Volume35, Issue4.Pages 1407-1417.  https://doi.org/10.1118/1.2842076
40. Andrew Whetten. Smoothing Splines of Apex Predator Movement: Functional modeling strategies for exploring animal behavior and social interactions. *Authorea*. September 24, 2021.

41. Torres LG, Thompson DR, Bearhop S, Votier SC, Taylor GA, Sagar PM, et al. White-capped albatrosses alter fine-scale foraging behavior patterns when associated with fishing vessels. Mar Ecol Prog Ser. 2011; 428: 289–301.
42. Gel'fand, I.M.; Yaglom, A.M. (1957). "Calculation of amount of information about a random function contained in another such function". American Mathematical Society Translations. Series 2. 12: 199–246. doi:10.1090/trans2/012/09. ISBN 9780821817124. English translation of original in Uspekhi Matematicheskikh Nauk 12 (1): 3-52.
43. Ramsay, J.O., and Silverman, B.W. (2005). Functional Data Analysis (second edition). Springer, New York.